\documentclass{aa}  
\usepackage{graphicx}
\usepackage{txfonts, color}

\newcommand{\msol}{\mbox{$M_{\odot}$}} 
\newcommand{\rsol}{\mbox{$R_{\odot}$}}

\newcommand{\lsol}{\mbox{$L_{\odot}$}}

\begin{document}

\title{The period -- luminosity and period -- radius relations of Type II and anomalous Cepheids 
in the Large and Small Magellanic Clouds}

%   \subtitle{}

   \author
   {
   M.~A.~T. Groenewegen\inst{1}
\and
   M.~I. Jurkovic\inst{2,3}
   }

   \institute{
              Koninklijke Sterrenwacht van Belgi\"e (KSB), 
              Ringlaan 3, B-1180 Brussels, Belgium \\
              \email{martin.groenewegen@oma.be}
         \and
              Astronomical Observatory of Belgrade (AOB),
              Volgina 7, 11 060 Belgrade, Serbia \\
              \email{mojur@aob.rs}
         \and     
			 Konkoly Observatory, Research Centre for Astronomy and Earth Sciences, 
			 Hungarian Academy of Sciences,
			 H-1121 Budapest, Konkoly Thege Mikl\'{o}s \'{u}t 15-17., Hungary \\
             }

   \date{Received ...; accepted ...}

% \abstract{}{}{}{}{} 
% 5 {} token are mandatory
 
  \abstract
  {Type II Cepheids (T2Cs) and anomalous Cepheids (ACs) are pulsating stars that follow separate period-luminosity relations.}
   {We study the period-luminosity ($PL$) and period-radius ($PR$) relations for T2Cs and ACs in the Magellanic Clouds.
   }
   {In an accompanying paper we determined luminosity and effective temperature for the 335 T2Cs and ACs in the 
LMC and SMC discovered in the OGLE-III survey, by constructing the spectral energy distribution (SED) and fitting 
this with model atmospheres and a dust radiative transfer model (in the case of dust excess). 
Building on these results we study the $PL$- and $PR$ relations.
Using existing pulsation models for RR Lyrae and classical Cepheids we derive the 
period-luminosity-mass-temperature-metallicity relations, and then estimate the pulsation mass.
	}
   {The $PL$ relation for the T2Cs does not appear to depend on metallicity, and, excluding the dusty RV Tau stars, is 
$M_{\rm bol}= +0.12 -1.78 \log P$ (for $P < 50$ days). Relations for fundamental and first overtone LMC ACs are also presented.
The $PR$ relation for T2C also shows little or no dependence on metallicity or period. 
Our preferred relation combines SMC and LMC stars and all T2C subclasses, and is
$\log R = 0.846  + 0.521 \log P$. Relations for fundamental and first overtone LMC ACs are also presented. 
The pulsation masses from the RR Lyrae and classical Cepheid pulsation models agree well for the short period T2Cs, 
the BL Her subtype, and ACs, and are consistent with estimates in the literature, i.e.  
$M_{\rm BLH} \sim 0.49$ \msol\ and $M_{\rm AC} ~\sim 1.3$ \msol, respectively. 
The masses of the W Vir appear similar to the BL Her. The situation for the pWVir and RV Tau stars is less clear. 
For many RV Tau the masses are in conflict with the standard picture of (single-star) post-AGB evolution, the
masses being either too large ($\gtrsim$ 1 \msol) or too small ($\lesssim$ 0.4 \msol).
	}
	{}

   \keywords{stars: variables: Cepheids: Type II Cepheids --- 
             stars: variables: Cepheids: Anomalous Cepheids --- 
             stars: fundamental parameters --- Magellanic Clouds ---
             stars: distances}

	\authorrunning{Groenewegen \& Jurkovic}
	\titlerunning{The $PL$ and $PR$ relations of T2Cs and ACs in the LMC and SMC}
   \maketitle
%
%________________________________________________________________

\section{Introduction}

Type II Cepheids (T2Cs) and anomalous Cepheids (ACs) are pulsating stars located in the instability strip (IS) of the 
Hertzsprung-Russell diagram (HRD) also occupied by the classical Cepheids (CCs) and RR Lyrae (RRL) variables. 
T2C are classically divided into subgroups based on their period, and, following \citet{Soszynski2008} 
and \citet{Soszynski2010}, they are the BL Herculis (BLH) ($1 - 4$ days), the (peculiar) W Virginis ((p)WVir) ($4 - 20$ days), 
and the RV Tauris (RVT) ($20 - 70$ days). The pulsation period of the ACs (from $\sim 0.9$ to $\sim 2$~days) overlaps 
with the short period T2Cs.
T2Cs are known to pulsate in fundamental mode (FU) only, while the ACs pulsate in the first overtone (FO) and fundamental mode (FU).

An important characteristic of T2Cs and ACs is that they follow a period-luminosity ($PL$) relation, and that these objects 
can be found in globular clusters, and galaxies \citep{Catelan_puls_book_2015}.
They can therefore be used in the calibration of the distance scale. T2Cs in particular are useful where CCs 
are too few and RRL too faint (see for example the review by \citet{Sandage_Tammann_2006} and \citet{Wallerstein_2002}).
ACs fill a space on the $PL$-diagram above the RRL and T2Cs by $\sim$0.5 to $\sim$2 magnitudes (as the period increases), 
but do not reach into the CC region. \citet{Caputo_2004} investigated the possibility that they do 
continue to the $PL$ relation of CCs, but \citet{Fiorentino_Monelli_2012} concluded that they are metal-poor stars 
that would not evolve into the CC IS region.

Period - luminosity relations have been discussed in several papers over the past decades.
\citet{Nemec_1994} is a rare example were $P-L-$[Fe/H] relations are derived for T2Cs and ACs (as well as RRL and SX Phe stars)
in $(B, V, K)$ colours based on objects in globular clusters. 
Recently, \citet{Clementini_GDR1} presented $PL$ relations for T2Cs and FU and FO ACs in the {\it Gaia} $G$-band.
\citet{Marconi_2004} provided a theoretical mass-dependent Period-Magnitude-Colour (PMC), Period-Wesenheit (PW) and 
Period-Magnitude-Amplitude relations for ACs in the metallicity range $Z= 0.0001 - 0.0004$. 
They also give the empirical $PW(VI)$ relation based on ACs observed in 7 dwarf spheroidal galaxies.
\citet{Ripepi_2014}, as a part of the VISTA Magellanic Cloud (VMC) Survey \citep{Cioni_VMC}, provided 
the $PL$ relation in the $K_{\rm s}$ band, and the $PW(V,K)$ relation for FU and FO ACs in the LMC, as well as the 
$PL$ relation in the $V$ and $I$ band, a $PMC$ and the $PW(V,I)$ relation based on the original OGLE-data.

\citet{DiCriscienzo_2007} derived theoretical period-magnitude ($PM$) in the NIR and period-Wesenheit ($PW$)
for various optical and NIR colour combinations relations combining pulsation models and evolutionary tracks 
for stars with periods up to 8 days, i.e. BLHs.
\cite{Matsunaga2011} present NIR $PL$ relations and Wesenheit relations for T2C in the SMC, and compare the 
results to their earlier work on the LMC \citep{Matsunaga2009} and Galactic Centre (GC) \citep{Matsunaga2006}.
The $K$-band $PL$ relation for T2C in the GC was also presented by \cite{GrUB08}, and the absolute
calibration was considered by \citet{Feast_2008}.
More recently, \citet{Manick_2017} used the OGLE-III LMC T2Cs to derive the Wesenheit $PW$ relation.

Recent survey work in the NIR allowed a reappraisal of the $PL$ relations, notably, \citet{Ripepi_2015} considered 
VMC data \citep{Cioni_VMC} to present several $PL$, $PLC$ and $PW$ relations, while
\citet{Bhardwaj2017} did a similar study using NIR data from the LMC Synoptic Survey \citep{Macri_2015}.

Apparently, there has been little recent work done on the radii of T2Cs.
\citet{BM1986} give a period-radius ($PR$) relation based on older data, and \citet{Balog_1997} derived the radii 
for 17 Galactic T2Cs using the Baade-Wesselink method. Figure~5 in their article displays a $PR$ relation, 
but they did not give an equation for their fit.

Masses for the T2Cs were estimated by \citet{Bono_1997_T2CEP} to be in a range between $0.52$ and $0.59$~\msol\ 
(for $Z <0.001$) for stars with periods below 15 days. Quoting the results of \citet{Vassiliadis_Wood_1993}, 
\citet{Wallerstein_2002} gives the initial mass of the brighter T2Cs (the RVT) to be around $1$~\msol. 
In the case of ACs pulsation models have been considered by various authors to find masses in the range 1.3-2.2~\msol\ 
(for $Z$ = 0.0001 and 0.0004, \citet{Bono_1997_ANCEP}, \citet{Marconi_2004}), or, specifically $1.2 \pm 0.2$~\msol\ for 
the ACs in the LMC \citep{Fiorentino_Monelli_2012}. Recently, \citet{MartinezVazques_2016} find $\sim$ 1.5~\msol\ for 
four ACs in the Sculptor dSph galaxy.

In the accompanying paper, (\citet{GJ2017a}, hereafter GJ17) studied all 335 T2C and AC in the 
Small and Large Magellanic Clouds (MCs) detected in the OGLE-{\sc III} data (\citet{Soszynski2008}, 
\citet{Soszynski2010}, \citet{Soszynski2010_ANC}\footnote{\citet{Soszynski2010_ANC} originally listed the 6 SMC ACs 
with a classical Cepheid identification number. In the OGLE-III Variable Stars 
Database (http://ogledb.astrouw.edu.pl/$\sim$ogle/CVS/ they 
were subsequently listed under the names that we use in GJ17 and the present paper, OGLE-SMC-ACEP 01...06.}).
The spectral energy distributions (SEDs) were constructed using
photometry from the literature and fitted with the dust radiative transfer code "More of DUSTY" (MoD, \cite{Gr_MOD}), 
an extension of the DUSTY radiative transfer code DUSTY \citep{Ivezic_D}.
Luminosities and effective temperatures were derived and are given in the Appendix in GJ17. 
The resulting Hertzsprung-Russell diagram was compared in a qualitative way to modern evolutionary tracks. 
In agreement with the findings cited above the BL Her can be explained by stars in the mass range 
$\sim 0.5-0.6$~\msol\ and the ACs by stars in the mass range $\sim 1.1-2.3$~\msol.
The origin of the (p)WVir is unclear however: tracks of  $\sim 2.5-4$~\msol\ cross the IS at the correct luminosity, as well as
(some) lower mass stars on the AGB that undergo a thermal pulse when the envelope mass is small, but the timescales make these
unlikely scenarios. 

For $\sim60\%$ of the RVT and $\sim 10\%$ of the W Vir (including the pWVir) objects an infrared excess was 
detected from the SED fitting. The results of \cite{Kamath2016} were confirmed that there exist stars with 
luminosities below that predicted from single-star evolution that show a clear infrared excess.
The light curves of more than 130 systems were investigated to look for the light-travel time (LTT) effect 
or light-time effect (LITE) \citep{Irwin52} in so-called {\it observed minus calculated}, (O-C)-diagrams. 
Twenty possible new binaries were identified, and about 40 stars that show a significant period change.

Previous work concentrated almost exclusively on deriving the $PL$ relation in the NIR bands or
using the Wesenheit index, the main aim of this paper is to use the 
stellar luminosity as parameter and in that way study the properties of these stars in a more fundamental way.
We used the results published in GJ17 to derive the period-luminosity and $PR$ relations of T2Cs and ACs. 
In addition, we derive estimates of the masses of these stars, based on theoretical pulsation models of RRL and CCs.

In Section~\ref{Sec:PL} we discuss the derived $PL$ relations for T2Cs and ACs.
In Section~\ref{Sec:PR} the period-radius relation is presented, and 
in Section~\ref{Sec:Mass} we estimate the masses. 
In the Section~\ref{Sec:Disc} we discuss and summarise our findings.

\section{Period -- Luminosity relation}
\label{Sec:PL}

\begin{figure}
\centering
\includegraphics[width=\hsize]{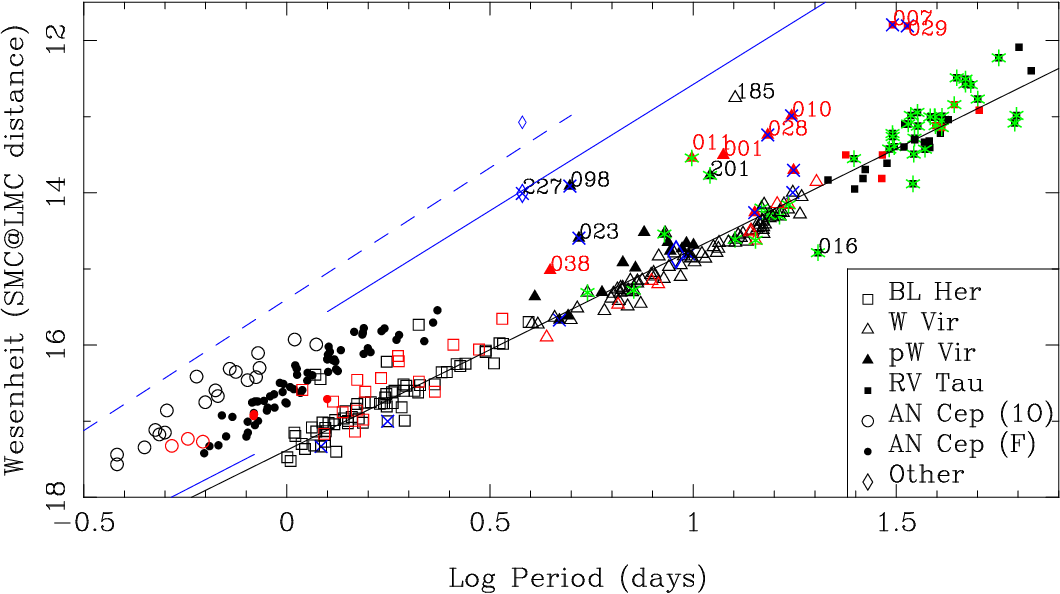}
\caption{
The Wesenheit $PL$ relation.
Stars in the SMC are plotted in red, and shifted to the distance of the LMC.
Some outliers are marked with their identifier.
Stars with an IR excess according to GJ17 are marked by a green plus sign.
Stars that show eclipsing or ellipsoidal variations according to OGLE are marked by a blue cross.
Stars plotted as a blue diamond are OGLE-LMC-CEP-0227 (the classical Cepheid in an eclipsing binary system, 
at $P= 3.79$~days), and the Galactic T2C $\kappa$ Pav (at $P= 9.08$~days), scaled to the distance of the LMC, see text.
For -0227 both the system value is plotted (the smaller, brighter point), and the Wesenheit magnitude 
of the Cepheid in the system (from \citet{Pilecki2013}).
The black solid line is the relation derived from the BLH + WVIR + non-dusty RVTs with periods below 50 days 
in the LMC (see Table~\ref{tablePL}), plotted over the entire period range.
The blue lines indicate the Wesenheit relation for CC in the LMC from \cite{Soszynski2008_DCEP} 
for FU (solid line) and FU (dashed line) pulsators, 
and the relation for RRab stars (at $\log P < -0.08$) from \cite{Soszynski2003}.
}
\label{figPLVI}
\end{figure}

Figure~\ref{figPLVI} first shows the classical $PL$ relation using the Wesenheit magnitude, 
$W= I - 1.55 \cdot (V-I)$, combining Fig.~1 in \cite{Soszynski2008} and Fig.~1 in \cite{Soszynski2010}.
In GJ17 we used distances to the LMC and SMC of 50 and 61~kpc in the SED modelling, and therefore we shifted 
the magnitudes of the SMC objects by 0.432 mag (distance moduli, DM, 18.927-18.495) to put them on the magnitude scale of the LMC.
The most prominent outliers are marked with their identifier.

Wesenheit $PL$ relations have been derived for various combinations of subclasses of T2Cs and ACs for both 
the LMC and SMC, and combined (``@LMC'', meaning the SMC objects have been placed at the distance of the LMC) 
and the results are listed in Table~\ref{tablePL}. Stars showing eclipsing or ellipsoidal variation 
(as identified by OGLE; the blue crosses in Figure~\ref{figPLVI}) have been excluded in the fitting 
and iterative 3$\sigma$ clipping has been applied to remove outliers.

Table~\ref{tablePL}, in addition, includes other determinations of the Wesenheit $PL$ relation from the literature, both 
observational as theoretical.
The Wesenheit $PL$ relation does not not seem to depend on metallicity. At characteristic periods of $\log P = 0.5$ (BLH), 
1.5 (RVT) and 1.0 (WVir, and the solutions that include BLH and/or RVT) the magnitude difference between the solutions for 
the SMC and LMC are within the errors consistent with the expected difference of 0.43 mag.

The derived relations are also in agreement with those listed in the literature although this is not so surprising as there 
are all based on the same OGLE-III data and only differ in details.
The RVT stars that show IR excess are brighter in $W$ than those without (also remarked by \citet{Manick_2017}), but 
excluding those there is well defined $W(VI)$ relation for BLH, WVIR and non-dusty RVT, as illustrated in Figure~\ref{figPLVI}.

The agreement with theoretical models is good for the BLH. The comparison with observations requires an adopted distance 
to the LMC, and an assumed metallicity for the models. For [Fe/H]$= -1$ and DM$= 18.50$ theory and observations 
agree within the error. The agreement is less good for the ACs, in particular the theoretical slope 
differs by almost 3$\sigma$ from the empirically derived one.

\begin{table*}
\setlength{\tabcolsep}{1.6mm}
\caption{Wesenheit and bolometric period-luminosity relations.} 
\label{tablePL}  

\centering                   
\begin{tabular}{lccrcccccr}     
\hline\hline 
Sample & Galaxy\tablefootmark{a} & Mag & Mag= \hfill $a$ & $+b \, \log P$    & dispersion & $\chi_r^2$\tablefootmark{b} & N &    N     & Ref. \\ 
       &        &     &                 &                   &   (mag)    &           &    & outliers &   \\
\hline        
BL Her    & LMC    & W & $17.359 \pm 0.022$  & $-2.576 \pm 0.080$ & 0.089 & 9.18 & 55 & 6 \\ 
BL Her    & SMC    & W & $17.558 \pm 0.134$  & $-2.429 \pm 0.480$ & 0.241 & 76.0 & 17 & 0 \\ 
BL Her    & @LMC   & W & $17.347 \pm 0.038$  & $-2.669 \pm 0.137$ & 0.170 & 32.8 & 74 & 4 \\  

 W Vir    & LMC    & W & $17.402 \pm 0.064$  & $-2.558 \pm 0.063$ & 0.093 & 9.91 & 76 & 2 \\ 
 W Vir    & SMC    & W & $18.329 \pm 0.168$  & $-3.009 \pm 0.158$ & 0.091 & 11.5 & 10 & 0 \\ 
 W Vir    & @LMC   & W & $17.471 \pm 0.061$  & $-2.624 \pm 0.060$ & 0.098 & 10.8 & 86 & 2 \\ 

RV Tau    & LMC    & W & $18.101 \pm 0.557$  & $-3.142 \pm 0.352$ & 0.249 & 72.6 & 41 & 1 \\ 
RV Tau    & SMC    & W & $17.707 \pm 1.049$  & $-2.585 \pm 0.674$ & 0.164 & 42.1 &  7 & 0 \\ 
RV Tau    & @LMC   & W & $18.004 \pm 0.494$  & $-3.077 \pm 0.313$ & 0.240 & 66.9 & 48 & 1 \\

BL Her + W Vir  & LMC  & W & $17.363 \pm 0.017$  & $-2.522 \pm 0.021$ & 0.102 & 11.6 & 133 &  6 \\ 
BL Her + W Vir  & SMC  & W & $17.597 \pm 0.072$  & $-2.356 \pm 0.103$ & 0.209 & 53.0 &  26 &  1 \\ 
BL Her + W Vir  & @LMC & W & $17.335 \pm 0.017$  & $-2.496 \pm 0.021$ & 0.108 & 13.2 & 153 & 13 \\ 

BL Her + W Vir + RV Tau\tablefootmark{c} & LMC  & W & $17.358 \pm 0.014$  & $-2.530 \pm 0.017$ & 0.089 & 9.00 & 136 & 10 \\ 
BL Her + W Vir + RV Tau\tablefootmark{c} & SMC  & W & $17.577 \pm 0.073$  & $-2.388 \pm 0.097$ & 0.232 & 64.9 &  28 &  0 \\ 
BL Her + W Vir + RV Tau\tablefootmark{c} & @LMC & W & $17.355 \pm 0.017$  & $-2.526 \pm 0.020$ & 0.118 & 15.8 & 162 & 12 \\

AC FU & LMC & W & $16.612 \pm 0.020$  & $-3.158 \pm 0.141$ & 0.150 & 25.7 &  62 &  0 \\ 
AC FO & LMC & W & $16.029 \pm 0.058$  & $-3.373 \pm 0.247$ & 0.140 & 24.2 &  19 &  0 \\

\\
% literature below

BL Her + W Vir  &  LMC & W & $17.364 \pm 0.015$ &  $-2.521 \pm 0.022$  & 0.105 & & 131 & & 1  \\  
BL Her + W Vir  &  SMC & W & $17.554 \pm 0.083$ &  $-2.304 \pm 0.107$  & 0.230 & & 27  & & 2 \\

BL Her + W Vir + RV Tau\tablefootmark{d} &  LMC   & W   & $17.33 \pm 0.03$ & $-2.53 \pm 0.03$ &  & & & & 3 \\ 

BL Her           & Theory & W & $17.30 \pm 0.07$ & $-2.43 \pm 0.02$ &    & & & & 4  \\ 

AC FU            & LMC & W   &  $16.59 \pm 0.02$ & $-3.41 \pm 0.16$ & 0.15 & & & & 5 \\
AC FO            & LMC & W   &  $16.05 \pm 0.05$ & $-3.44 \pm 0.22$ & 0.13 & & & & 5 \\

AC FU            & Theory & W   &  $16.55 $ & $-2.94 $ &  & & & & 6 \\

\\

BL Her          &  LMC & $M_{\rm bol}$  & $+0.141 \pm 0.051$  & $-1.749 \pm 0.200$ & 0.274 & 33.0 &  57 &  4 \\ 
BL Her          &  SMC & $M_{\rm bol}$  & $-0.250 \pm 0.176$  & $-0.691 \pm 0.717$ & 0.302 & 64.9 &  15 &  2 \\
BL Her          &  MCs & $M_{\rm bol}$  & $-0.027 \pm 0.065$  & $-1.326 \pm 0.257$ & 0.282 & 89.5 &  72 &  6 \\

W Vir           &  LMC & $M_{\rm bol}$  & $0.723 \pm 0.115$   & $-2.358 \pm 0.119$ & 0.186 & 36.8 &  74 &  5 \\
W Vir           &  SMC & $M_{\rm bol}$  & $0.965 \pm 0.318$   & $-2.589 \pm 0.319$ & 0.210 & 33.2 &  10 &  0 \\
W Vir           &  MCs & $M_{\rm bol}$  & $0.743 \pm 0.109$   & $-2.379 \pm 0.112$ & 0.201 & 37.1 &  85 &  4 \\

RV Tau\tablefootmark{d} &  LMC & $M_{\rm bol}$  & $+1.442 \pm 1.146$   & $-2.919 \pm 0.750$ & 0.301 & 91.1 &  15 &  0 \\
RV Tau\tablefootmark{d} &  SMC & $M_{\rm bol}$  & $-1.088 \pm 0.433$   & $-1.367 \pm 0.290$ & 0.041 &  4.6 &   4 &  0 \\
RV Tau\tablefootmark{d} &  MCs & $M_{\rm bol}$  & $+0.951 \pm 0.974$   & $-2.620 \pm 0.639$ & 0.298 & 78.4 &  19 &  0 \\

BL Her + W Vir  &  LMC & $M_{\rm bol}$ & $+0.199 \pm 0.035$  & $-1.827 \pm 0.042$ & 0.230 & 40.6 & 130 & 10 \\
BL Her + W Vir  &  SMC & $M_{\rm bol}$ & $-0.087 \pm 0.100$  & $-1.561 \pm 0.182$ & 0.349 & 256. &  26 &  1 \\
BL Her + W Vir  &  MCs & $M_{\rm bol}$ & $+0.068 \pm 0.037$  & $-1.704 \pm 0.049$ & 0.267 & 83.1 & 159 &  8 \\

BL Her + W Vir + RV Tau\tablefootmark{c}   & LMC & $M_{\rm bol}$ & $+0.226 \pm 0.033$ & $-1.870 \pm 0.039$ & 0.233 & 40.5 & 136 & 11 \\
BL Her + W Vir + RV Tau\tablefootmark{c}   & SMC & $M_{\rm bol}$ & $-0.048 \pm 0.101$ & $-1.686 \pm 0.172$ & 0.370 & 275. &  27 &  1 \\
BL Her + W Vir + RV Tau\tablefootmark{c, e} & MCs & $M_{\rm bol}$ & $+0.119 \pm 0.036$ & $-1.787 \pm 0.044$ & 0.276 & 81.8 & 166 &  9 \\

AC FU           &  LMC & $M_{\rm bol}$  & $-0.436 \pm 0.033$   & $-3.122 \pm 0.213$ & 0.255 & 71.3 &  61 &  1 \\
AC FO           &  LMC & $M_{\rm bol}$  & $-1.126 \pm 0.074$   & $-3.248 \pm 0.305$ & 0.244 & 53.2 &  20 &  0 \\

\hline       
\end{tabular}
\tablebib{
(1)~\citet{Matsunaga2009}; (2)~\citet{Matsunaga2011}; (3)~\citet{Manick_2017}; 
(4)~\citet{DiCriscienzo_2007} for [Fe/H] $= -1$, $l/H_{\rm p}= 1.5$, and LMC distance modulus 18.50; 
(5)~\citet{Ripepi_2014} ; (6)~\citet{Marconi_2004} for $M= 1.3$ \msol and LMC distance modulus 18.50.
}
\tablefoot{
\tablefoottext{a}{For the $PL$ relations in the Wesenheit index ``@LMC'' means the stars in the LMC plus 
the stars in the SMC placed at the distance of the LMC by a shift of $0.432$ magnitude.}
\tablefoottext{b}{The reduced $\chi^2$ is based on an assumed ``error'' in the Wesenheit index and 
bolometric magnitude of $0.03$ mag.}
\tablefoottext{c}{Excluding RVTs with dust excess and for $P < 50$~days.}
\tablefoottext{d}{Excluding RVTs with dust excess.}
\tablefoottext{e}{The preferred solution.}
}
\end{table*}

\begin{figure}
\centering
\includegraphics[width=\hsize]{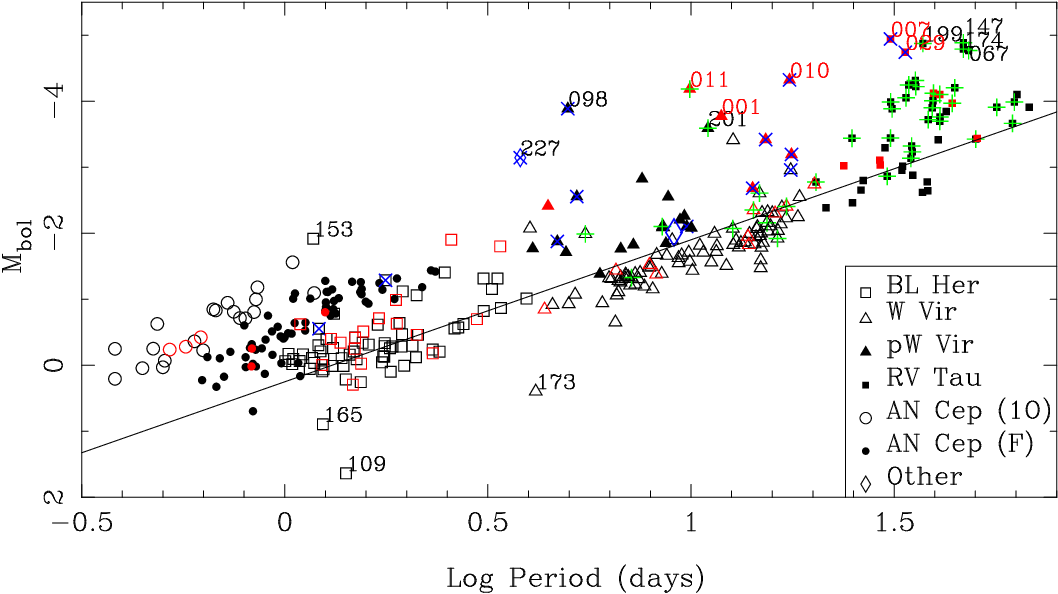}
\caption{
Bolometric $PL-$relation.
Stars in the SMC are plotted in red.
The error in $M_{\rm bol}$ is smaller than the plot symbol.
Some outliers are plotted with their identifier.
Stars with an IR excess according to GJ17 are marked by a green plus sign.
Stars that show eclipsing or ellipsoidal variations according to OGLE are  marked by a blue cross.
The classical Cepheid in the eclipsing binary OGLE-LMC-CEP-0227 (at $P= 3.79$~days) and the Galactic T2C $\kappa$ Pav 
(at $P= 9.08$~days) are plotted as blue diamonds.
The black solid line is the relation derived from the BLH + WVIR + non-dusty RVTs with periods below 50 days 
in the LMC (see Table~\ref{tablePL}), plotted over the entire period range.
}
\label{figPL}
\end{figure}

Figure~\ref{figPL} shows the bolometric version of the $PL$ relation, using the luminosities derived in GJ17.
The bottom part of Table~\ref{tablePL} gives the corresponding fits to the $PL$ relation.
What is immediately noticeable is that the scatter in the bolometric $PL$ relations is significantly larger than in the 
corresponding Wesenheit relations.
There could be several reasons for this. First, the Wesenheit relations are based on two intensity-mean magnitudes, while 
the luminosities are derived based on a fit to the entire SED, that is based on non-contemporaneous photometry. 
Second, if there are issues related to blending or binarity then certain combinations of the parameters involved may still 
yield a Wesenheit index close to the mean relation, but the fitting of the entire SED will more likely yield deviant results.

For comparison we will refer to two other systems throughout the discussion, namely the best studied of the known CCs in 
an eclipsing binary in the LMC (OGLE-LMC-CEP-0227), and one of the best studied T2C in our Galaxy, $\kappa$ Pav 
(WVir type, $P=9.09$~days). 
The latter has a HST-based distance of $180 \pm 9$ pc \citep{Benedict_2011} and a metallicity of [Fe/H]= 0.0 \citep{LB1989}. 
It is not listed in the 1st {\it Gaia} data release \citep{GAIAC1}. Time-series photometry in $V, I$ is available 
from \citet{Berdnikov2008}, from which we derived the mean magnitudes. In Figure~\ref{figPLVI} it is plotted as 
if it were located at the adopted distance to the LMC. 
\citet{Breitfelder_2015} quote an effective temperature of $T_{\rm eff}= 5739 \pm 107$~K, implying $L= 508 \pm 65$~\lsol.
\citet{Pilecki2013} have derived the $V, I$ magnitudes of the two components in OGLE-LMC-CEP-0227, and the Wesenheit magnitude 
of the Cepheid in the system and of the total binary system are plotted in Figure~\ref{figPLVI} (at $P=$ 3.79d), together 
with the Wesenheit relation for FU and FO CCs in the LMC from \cite{Soszynski2008_DCEP}. \citet{Pilecki2013} also 
derived $\log L = 3.158 \pm 0.049$~\lsol and $T_{\rm eff}= 6050 \pm 160$~K, in agreement with \citet{Marconi_2013} 
($\log L = 3.16 \pm 0.02$~\lsol, $T_{\rm eff}= 6100 \pm 50$~K).

$\kappa$ Pav, that is close to the estimated $PL$ relation in the Wesenheit index, is brighter than the relation in 
bolometric magnitude. The derived radius and effective temperature \citep{Breitfelder_2015} imply 
$M_{\rm bol}= -2.01 \pm 0.13$ while the various $PL$ relations gives values in the range $-1.53$ to $-1.59$. 
This could be due to the intrinsic width of the IS, the assumed distance (although the HST-based distance is 
accurate to 5\%), a metallicity dependence of the $PL$ relation (this is not obvious from a comparison of 
LMC and SMC objects), or the fact that it has a binary companion. 
In fact, \citet{Matsunaga2009} discusses $\kappa$ Pav in detail, and suggests that it should be classified as a pWVir object.

The outliers that are marked in both figures are mostly pWVir type stars, some of which have been classified as binaries 
by the OGLE team: OGLE-LMC-T2CEP-098, and -023, OGLE-SMC-T2CEP-007, -010, -028, or where the LITE was tentatively 
detected in GJ17, OGLE-SMC-T2CEP-001, and -029. The other outliers are mostly dusty objects with infrared excess detected 
in GJ17 namely some pWVir objects (OGLE-LMC-T2CEP-201 and OGLE-SMC-T2CEP-011) and mostly 
RVT (OGLE-LMC-T2CEP-016, -067, -147, -174, -199).

\section{Period -- Radius relation}
\label{Sec:PR}

\begin{figure}
\centering
\includegraphics[width=\hsize]{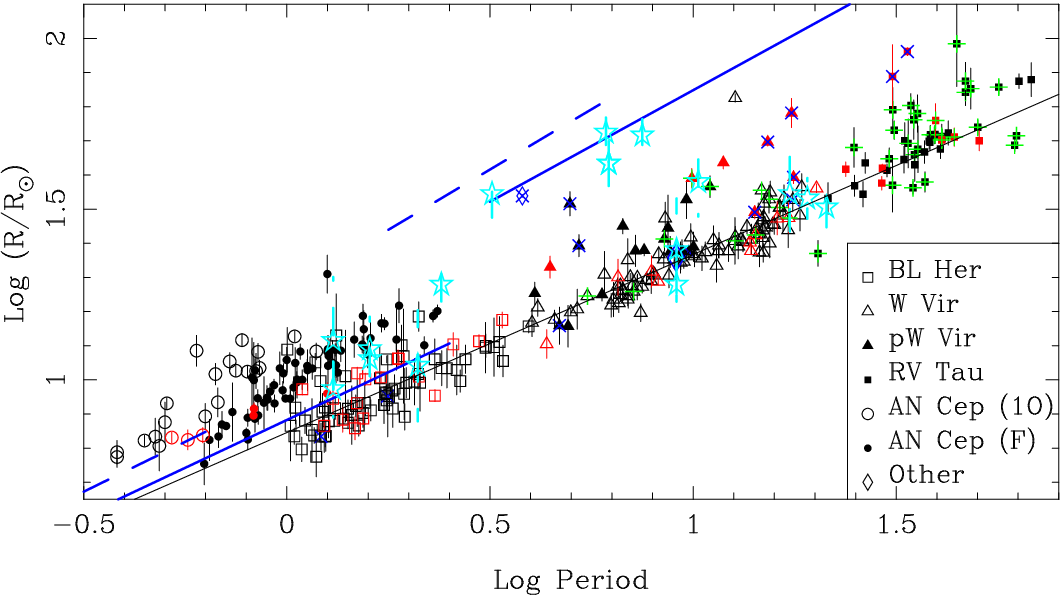}
\caption{
The Period-radius relation.
Stars in the SMC are plotted in red.
The black solid line is a fit to the BL Her, W Vir and RV Tau stars in the LMC (excluding 3$\sigma$ outliers).
The blue solid and dashed lines are the theoretical $PR$ relations for FU and FO RRL (shorter periods) 
and CC in the LMC, respectively (see text).
The values for the Cepheid in the eclipsing binary system OGLE-LMC-CEP-0227 (at $P= 3.79$~days) and the 
Galactic T2C $\kappa$ Pav (at $P= 9.08$~days) are plotted as blue diamonds. 
Light blue stars are represent the stars from \citet{Balog_1997}. Details about these stars are given in the text.
}
\label{figPR}
\end{figure}

Figure~\ref{figPR} shows the $PR$ relation based on the derived effective temperatures and luminosities in GJ17. 
The resulting radii with error bars are given in Table~\ref{App:Mass}.

$PR$ relations have been derived for various combinations of the T2C subclasses in the SMC, LMC, and both, and for the 
FU and FO ACs in the LMC and the results are listed in Tab.~\ref{tablePR}.
Stars showing eclipsing or ellipsoidal variation have been excluded in the fitting and iterative 3$\sigma$ clipping has 
been applied to remove outliers. Contrary to the $PL$ relations where the dusty RVT stars deviated significantly and where 
excluded this is not the case here.

\citet{Marconi_2015} present the latest nonlinear, time-dependent convective hydrodynamical models of RRL stars 
for different metallicites and masses. Specifically they present Period-Mass-Radius-Metallicity (PMRZ) relations for 
fundamental and first-overtone pulsators (Their Eqs.~7 and 8). As they were concerned with RRL they 
excluded "the sequence D models" (see \citet{Marconi_2015} for details) in their fitting procedure, 
since these luminosity levels were considered too bright for typical RRLs.
However, these luminosities are typical for T2C, and therefore we re-derived the PMRZ relation 
for all models with $\log L > 1.65$~\lsol\ (and that reach up to $\log L \sim 2.0$, and periods up to $\sim 2.4$ days) 
using their dataset. We find: 
\begin{align}
\log R = & \; (0.763 \pm 0.003) -(0.037 \pm 0.001) \log Z  \nonumber \\
         & \; \qquad {} \qquad {} \qquad {} \;  +(0.560 \pm 0.004) \log P \;\; {\rm (N= 195)}
\end{align}
for FU pulsators, and
\begin{align}
\log R= & \; (0.855 \pm 0.005) -(0.034 \pm 0.001) \log Z  \nonumber \\
        & \; \qquad {} \qquad {} \qquad {} \; +(0.585 \pm 0.007) \log P \;\; {\rm (N= 63)}
\end{align}
for FO pulsators.
These relations are plotted in Fig.~\ref{figPR} at the average metallicity of 
RRL in the LMC \citep{Gratton_2004} of [Fe/H] $= -1.5$ (or $\log Z= -3.23$).
The theoretical relation lies above the observed one. The slope agrees within the error bar with 
the observed relation for BLH (see Table.~\ref{tablePR}), but the zero point is slightly larger. 

In a similar way, \cite{Bono_2000} present non-linear pulsation models for CCs for various masses and metallicities. 
Period-Radius relations for FU pulsators at three different metallicities were already presented in \citet{Bono_1998}.
Here we have re-derived the PMRZ relations for FU and FO pulsators from the \cite{Bono_2000} dataset, combining 
the "canonical" and "non-canonical" models (like they did), and find
\begin{align}
\log R= & \; (1.115 \pm 0.012) -(0.039 \pm 0.005) \log Z  \nonumber \\
        & \; \qquad {} \qquad {} \qquad {} \;  +(0.653 \pm 0.003) \log P \;\; {\rm (N= 202)}
\end{align}
for FU pulsators, and
\begin{align}
\log R= & \; (1.257 \pm 0.028) -(0.003 \pm 0.014) \log Z  \nonumber \\
        & \; \qquad {} \qquad {} \qquad {} \; +(0.706 \pm 0.016) \log P \;\; {\rm (N= 27)}
\end{align}
for FO pulsators.
These relations are plotted in Fig.~\ref{figPR} at the average metallicity of 
Cepheids in the LMC \citep{Romaniello_2008} of [Fe/H] $= -0.33$ (or $\log Z= -2.06$).

For comparison we have plotted objects with known radii. 
The values for the Cepheid in OGLE-LMC-CEP-0227 and Galactic T2C $\kappa$ Pav are plotted as blue diamonds. 
Based on a Baade-Wesselink type analysis. \citet{Breitfelder_2015} derived a projection factor of $p= 1.26 \pm 0.04 \pm 0.06$, 
and a radius of $R= 22.83 \pm 1.14$ \rsol.
Big light-blue stars represent Galactic Type II Cepheids that had their radii derived by \citet{Balog_1997} using the 
Baade-Wesselink method. In case of $\kappa$ Pav ($R = 19 \pm 5$~\rsol) their result is in good agreement 
with \citet{Breitfelder_2015}. 
It must be noted that since  the article was published in 1997, some of the objects have been reclassified. That explains why 
they scatter so much. Looking at the classification by the General Catalog of Variable Stars 
(GCVS)\footnote{http://www.sai.msu.su/gcvs/gcvs/} and the International Variable Star Index 
(VSX)\footnote{https://www.aavso.org/vsx/} it appears that KL Aql (P= 6.1 day), V733 Aql (P= 6.2), BB Her (P= 7.5) 
and DR Cep (P= 19.1) are CCs. DQ And (P= 3.2) has a questionable classification, but it is more likely than not that is a CCs, too. 
The cases of TX Del (P= 6.2) and IX Cas (P= 9.1) are different, because their radius from the Baade-Wesselink analysis might 
have been influenced with the fact that they are in binary systems.  AU Peg (P= 2.4) is hard to interpret, because it was 
suggested that it might not be a T2C, and the radius of $19 \pm 4$~\rsol\ puts it above the $PR$ relation for T2Cs. 
It is also a binary, so, again, that could have had an influence on the determined radius. 
BL Her (P= 7.5), XX Vir (P= 1.3), SW Tau (P=1.6), NW Lyr (P=1.6), V553 Cen (P= 2.1, a C-rich object), $\kappa$ Pav, AL Vir (P= 10.3), 
W Vir (P= 17.3) and V1181 Sgr (P= 21.3) are T2Cs, and they follow our $PR$ relation.

There is no obvious dependence of the $PR$ relation on metallicity, or on subclass.
Within the error bars, all T2C can be represented by a single $PR$ relation (solutions 13-15). For this type of relation the slope 
between the solution for SMC and LMC stars differ by 2$\sigma$, but slope and zero point are not independent. 
At the characteristic period of 10 days the predicted radii for an SMC and LMC T2C are identical. 
Combining both galaxies and all periods, solution (15) becomes our preferred $PR$ relation for T2Cs.

Table~\ref{tablePR} also includes the old relation presented in \citet{BM1986}. 
They did not give error bars, but at face value the relation is similar to the preferred relation we derive for the MC T2Cs.
Together with the T2C from \citet{Balog_1997} this supports the suggestion that there is no strong dependence of the 
$PR$ relation on metallicity.

\begin{table*}
\caption{Period-Radius relations.} 
\label{tablePR}  

\centering                   
\begin{tabular}{lcccccrccl}     
\hline\hline 
Sample & Galaxy   & $\log R=$ \hfill $a$ & $+b \, \log P$  & dispersion & $\chi_r^2$ & N & N       & Solution & Ref. \\ 
       &          &  (\rsol)             &                 &            &           &   & outliers &       &   \\ 
\hline        
BL Her & LMC      & $0.830 \pm 0.013$  & $0.564 \pm 0.049$ & 0.047 & 2.60 & 57 & 4 & (1) \\   
BL Her & SMC      & $0.852 \pm 0.028$  & $0.574 \pm 0.117$ & 0.056 & 12.4 & 17 & 0  & (2)\\   
BL Her & MCs      & $0.847 \pm 0.013$  & $0.551 \pm 0.052$ & 0.058 & 5.43 & 76 & 2 & (3) \\   

 W Vir & LMC      & $0.823 \pm 0.020$  & $0.541 \pm 0.021$ & 0.037 & 1.74 & 77 & 2 & (4) \\   
 W Vir & SMC      & $0.709 \pm 0.079$  & $0.620 \pm 0.071$ & 0.038 & 4.11 & 10 & 0 & (5) \\   
 W Vir & MCs      & $0.828 \pm 0.020$  & $0.531 \pm 0.020$ & 0.037 & 2.25 & 87 & 2 & (6) \\   

RV Tau & LMC      & $0.848 \pm 0.141$  & $0.528 \pm 0.088$ & 0.076 & 5.72 & 41 & 1 & (7) \\   
RV Tau & SMC      & $0.977 \pm 0.188$  & $0.440 \pm 0.124$ & 0.039 & 2.54 &  7 & 0 & (8) \\   
RV Tau & MCs      & $0.864 \pm 0.112$  & $0.517 \pm 0.071$ & 0.072 & 5.16 & 48 & 1 & (9) \\   

BL Her + W Vir & LMC  & $0.837 \pm 0.007$  & $0.528 \pm 0.008$ & 0.041 & 2.09 & 134 & 6 & (10) \\
BL Her + W Vir & SMC  & $0.869 \pm 0.015$  & $0.480 \pm 0.022$ & 0.050 & 9.81 &  27 & 0  & (11) \\
BL Her + W Vir & MCs  & $0.852 \pm 0.006$  & $0.508 \pm 0.008$ & 0.044 & 3.46 & 161 & 6  & (12) \\

BL Her + W Vir + RV Tau & LMC  & $0.833 \pm 0.007$  & $0.535 \pm 0.007$ & 0.050 & 2.82 & 174 & 8  & (13) \\ 
BL Her + W Vir + RV Tau & SMC  & $0.861 \pm 0.013$  & $0.501 \pm 0.016$ & 0.050 & 8.75 &  34 & 0  & (14) \\
BL Her + W Vir + RV Tau\tablefootmark{a} & MCs  & $0.846 \pm 0.006$  & $0.521 \pm 0.006$ & 0.053 & 3.91 & 209 & 7  & (15) \\

AC FU   & LMC      & $0.972 \pm 0.005$  & $0.692 \pm 0.034$ & 0.045 & 2.39 & 61 & 1  & (16) \\   
AC FO   & LMC      & $1.113 \pm 0.016$  & $0.733 \pm 0.073$ & 0.054 & 3.65 & 20 & 0  & (17) \\   

T2C     & Galactic & 0.87 & 0.54 &                                  &     &    &     &      & (1) \\

\hline       
\end{tabular}
\tablebib{
(1)~\citet{BM1986}.
}
\tablefoot{
\tablefoottext{a}{The preferred solution.}
}
\end{table*}

\section{Masses from evolutionary models}
\label{Sec:Mass}

Following Section~\ref{Sec:PR} we derived the following equation for FU pulsators 
from the models in \citet{Marconi_2015} with $\log L > 1.65$,
\begin{align}
\log P = & \;  (11.468 \pm 0.049) +(0.8627 \pm 0.0028) \log L               \nonumber \\ 
         & \; -(0.617  \pm 0.015)  \log M -(3.463 \pm 0.012) \log T_{\rm eff} \nonumber \\
         & \; +(0.0207 \pm 0.0013) \log Z \;\; {\rm (N= 195, \sigma = 0.0044).}
\label{EqPRRL}
\end{align}
Similarly, we used the models in \cite{Bono_2000} to find for FU pulsators (cf. their Table~6),
\begin{align}
\log P = & \;  (10.649 \pm 0.085) +(0.9325 \pm 0.0053) \log L               \nonumber \\ 
         & \; -(0.799  \pm 0.020)  \log M -(3.282 \pm 0.022) \log T_{\rm eff} \nonumber \\
         & \; +(0.0393 \pm 0.0026) \log Z \;\; {\rm (N= 202, \sigma = 0.0085)}
\label{EqPCEP}
\end{align}
Equations~\ref{EqPRRL} and \ref{EqPCEP} allow us to derive the current mass if we know 
period, luminosity, effective temperature and metallicity.
The procedure seems to give sensible results. For the stellar parameters derived 
for the CC -0227 \citep{Pilecki2013}, and its metallicity of $Z= 0.004$ \citep{Marconi_2013}, 
the equation based on the Cepheid models gives a mass of $4.41 \pm 0.44$ \msol, 
in agreement with the masses found by \citet{Pilecki2013} ($4.165 \pm 0.032$ \msol) and 
\citet{Marconi_2013} ($4.14 \pm 0.06$ \msol).
However, the equation from the RRL pulsation modes also gives an estimate 
that is correct within the error bar, namely $5.86 \pm 1.18$ \msol.

We used Equations~\ref{EqPRRL} and \ref{EqPCEP} with $Z= 0.004$ (for both LMC and SMC) to estimate the masses.
For the overtone AC the equations were evaluated at their fundamental period $\log P_{\rm FU}= \log P_{\rm FO} +0.127$.
Errors were estimated from the error in $T_{\rm eff}$ and $L$, and the intrinsic scatter in the relation.
The results are given in Table~\ref{table:mass}.
The Cepheid and RRL-based masses agree within 3$\times$ the combined error bars (marked as ``OK'' in Table~\ref{table:mass})
or within 15\% irrespective of the error bars (marked as ``ok'' in Table~\ref{table:mass})
%   ``OK + ok''  (incl binaries)
in 90\% for the BLH (72/80), 82\% for the FU ACs (53/65), 74\% for the FO ACs (17/23), 
58\% for the pWVir (17/24), but only 22\% for the WVir (20/90) and 18\% (9/51) for the RVT.

The effect of changing the metallicity was investigated.  As a test it was lowered 
to $Z= 0.0012$ that would give a mass for the Cepheid -0227 from the Cepheid models 
in agreement with the observations (see above). 
This improved the agreement between the two mass estimates for the BLH and WVir, 
slightly reduced the agreement for the FU ACs and left the other percentages unchanged.

For the estimates based on $Z= 0.004$ the geometric mean of the two estimates was taken. 
For the various pulsation classes we find the following range in masses 
(listed are the 10,50,90\% percentiles), where known eclipsing and ellipsoidal variables were excluded: 
BL Her (0.36, 0.49, 0.87~\msol),
W Vir (0.31, 0.41, 0.57~\msol),
pW Vir (0.37, 0.74, 1.29~\msol),
RV Tau (0.25, 0.43, 0.82~\msol), and   
ACs (0.89, 1.29, 1.90~\msol), with the same range for FU and FO pulsators.
Assuming $Z= 0.0012$ would lower these mass estimates by about 5\%.
For the RVT it does not matter significantly whether one separates them into the dusty or non-dusty ones.

Taking only the stars where the two mass estimates agree (the ``OK'' and ``ok'' from Table~\ref{table:mass})
leaves these ranges essentially unchanged for the BLH and the ACs.
For the WVir the range becomes (0.43, 0.50, 0.66~\msol\ based on 20 stars).
The pWVir have been suggested to be in binary systems. Including eclipsing/ellipsoidal stars, and 
taking the stars where the two mass estimates agree the mass range is increased significantly to 
(0.72, 1.22, 1.77~\msol, for 14 stars).
The number of RVT where the two mass estimates agree (and are non-eclipsing/ellipsoidal) is only 7, 
and the median mass is 0.82~\msol.

The classes of objects where the mass estimates agree best and most are the BLH and ACs.
For these classes the estimates also agree with previous estimates in the literature.
For the WVir the situation is slightly less clear but the mass estimates are similar to those of the BLH.
The most confusing picture is presented by the pWVir and the RVT. 
The mass estimate for the former classes is definitely larger than for the BLH and WVir.

Some of the known binaries have a (spurious) large mass assigned: LMC -098 (3~\msol, pWVir), 
SMC -007 (1.9~\msol, RVT), -010 (2.3~\msol, pWVir), -028 (1.6~\msol, pWVir), and \mbox{-029} (2.5~\msol, RVT).
Based on this, the following stars (non ACs) could also be binaries:
The LMC objects
-032 (RVT,   1.9 \msol), 
-123 (BLHer, 2.2 \msol), 
-136 (BLHer, 2.0 \msol), 
-153 (BLHer, 1.6 \msol), 
-185 (WVir,  4.5 \msol), 
and SMC objects -001 (pWVir, 1.7 \msol), and   
-011 (pWVir, 1.8 \msol). 
The first 5 stars listed were also removed as outliers in the $PR$ relation, 
and -153 and -185 were also removed as outliers in the $P-M_{\rm bol}$ relation.
-185 was an outlier in the amplitude-magnitude diagram (Fig.~10 in GJ17).
None of these stars were marked as possible binaries based on the LTT effect in GJ17 however. 
-153 has been indicated by the OGLE team as blended, and the finding chart for -156 on the OGLE-III Variable Stars 
Database (http://ogledb.astrouw.edu.pl/$\sim$ogle/CVS/) suggests it is blended as well, so for those stars that 
could easily be the cause of their brighter appearance. 
On the other hand, based on the mass estimate, period, and the fact that it is brighter and larger than the other BLH, 
-123 could be classified as an FU AC. 

During the refereeing process \citet{Pilecki2017} appeared that analysed this system in detail and derived a mass of 
1.51 $\pm$ 0.09~\msol, $T_{\rm eff}= 5300 \pm 100$~K, and $L =  450 \pm   40$~\lsol\ for the pulsating star, and
6.8  $\pm$ 0.4~\msol,  $T_{\rm eff}= 9500 \pm 500$~K, and $L = 5000 \pm 1100$~\lsol\ for the secondary.
Fitting the SED as a single object GJ17 found $T_{\rm eff}= 7375 \pm 312$~K and $L= 2857 \pm  169$~\lsol.
Using the luminosity and effective temperature (and errors) from \citet{Pilecki2017} the mass estimate based 
on the Cepheid, respectively, RRL pulsation models is 1.37 $\pm$ 0.04~\msol, respectively, 1.52 $\pm$ 0.07~\msol,
in agreement with \citet{Pilecki2017}.
The parameters they derived, in particular the mass, depend on the adopted, so-called, projection factor, $p$.
We find that the geometric mean of the Cepheid and RRL pulsation mass, and their derived mass agrees 
best for $p= 1.32 \pm 0.03$, in excellent agreement with their adopted $p=1.30 \pm 0.04$.
Interestingly, their derived values of $M_{\rm bol}= -1.88 \pm 0.09$ and $R= 25.2 \pm 0.4$~\rsol\ 
(and that do not depend very much on the adopted value of $p$) still make the Cepheid
overluminous and oversized with respect to our preferred solutions of Tables~1 and 2 that give 
$M_{\rm bol}= -1.12 \pm 0.05$, and $R= 16.2 \pm 0.3$~\rsol.

The mass estimates for the RVT show both very large and very small values. 
As indicated above, many RVT have mass estimates that are well above that expected for a post-AGB object (0.55 - 1.1~\msol).
In addition, of the about 30 stars that have a mass estimate below 0.35~\msol, 10 are RVT and 6 of those have a dust excess.
Such low masses are also not expected from single-star evolution and, as remarked in in GJ17, the shape of the dust excess 
in these SEDs points to a disk-like structure that is thought to result from binary evolution.
Possibly some are related to the so called binary evolutionary pulsators (BEP), binary stars that appear in 
the IS after significant mass transfer. Recently, \citet{Karczmarek2016} did extensive simulations to find 
contaminations of genuine RRL and classical Cepheids of respectively, 0.8 and 5\% by BEP. In GJ17 we estimated that 
a contamination of several percent is plausible for T2C as well.

\section{Summary and conclusions}
\label{Sec:Disc}

The luminosities and effective temperatures derived in GJ17 for 335 T2Cs and ACs in the SMC and LMC were used to
study the period-Wesenheit, and for the first time to our knowledge in the 21st Century, the period-bolometric luminosity 
and $PR$ relations for these classes of stars.

The $P-M_{\rm bol}$ relation shows more scatter than the $PW$ relation. This is likely due to the fact that the 
fits to the SEDs presented in GJ17 are based on non-contemporaneous photometry over a large wavelength region. 
This will introduce some ``natural scatter'', but will likely reveal the effect of binarity or blending more easily
than when using only the OGLE mean $V,I$ magnitudes.
The  period-bolometric luminosity and $PR$ relations do not significantly depend on metallicity (as probed
by the T2C in the SMC and LMC, and supported by the limited data for Galactic T2Cs), and excluding 
the dusty RVTs, the T2C can be described by single relationships.

We used the published results of theoretical pulsation models for classical Cepheids and RRL 
to derive period-luminosity-mass-temperature-metallicity relations. Assuming a metallicity these relations allow us 
to derive the pulsation mass for all objects based on both types of models.
The RRL and CC models agree well and most for the BLH and ACs. The masses agree with those in the literature, 
respectively, $\sim0.5~\msol$ and $\sim1.3~\msol$.
For the RVT the agreement with the two mass estimates is poorest, and often indicates masses that are inconsistent with
single-star evolution of a post-AGB star: either above $\sim1~\msol$ or well below $\sim0.5~\msol$.

\begin{acknowledgements}
M.I.J. acknowledges financial support from the Ministry of 
Education, Science and Technological Development of the 
Republic of Serbia through the project 176004, and the 
Hungarian National Research, Development and Innovation Office through NKFIH K-115709.
This research has made use of the VizieR catalogue access tool, CDS,
Strasbourg, France. The original description of the VizieR service was
published in A\&AS 143, 23.

\end{acknowledgements}

%-------------------------------------------------------------------

\bibliographystyle{aa.bst}
	\bibliography{references.bib}

%-------------------------------------------------------------------
%\Online

\begin{appendix} %First online appendix

\section{Mass estimates}
\label{App:Mass}

Table~\ref{table:mass} contains the radii with error bars, and the mass estimates and error bars for individual object 
based on the Cepheid and RRL pulsation models (Eqs.~\ref{EqPRRL} and \ref{EqPCEP}). For reference it is also indicated 
if the star has an IR excess (Dusty=1), or is a known binary (Binary=1).
If the two mass estimates agree within 3$\times$ the combined error bars the last column has an "OK" listed, and
else, if the two mass estimates still agree within 15\% (without considering the errors) 
the last column has an "ok" listed.

\begin{table*}
\caption{Mass estimates}
\label{table:mass}

\begin{tabular}{lrrrrrrrrlrrrrr}
\hline
Name               & Type  & Period & Radius   & Mass$_{\rm Cep}$  & Mass$_{\rm RRL}$    & Dusty? & Binary? & Agree? \\ 
                   &       &  (d)   &  (\rsol) &     (\msol)     &     (\msol)       &        &         &            \\
\hline

OGLE-LMC-ACEP-001 & F     &   0.85 &  7.86 $\pm$  0.48 & 0.891 $\pm$ 0.016 & 1.023 $\pm$ 0.029 & 0 & 0 &  ok  \\ 
OGLE-LMC-ACEP-002 & F     &   0.98 &  9.32 $\pm$  0.56 & 1.127 $\pm$ 0.022 & 1.314 $\pm$ 0.045 & 0 & 0 &  ok  \\ 
OGLE-LMC-ACEP-003 & 1O    &   0.51 &  5.94 $\pm$  0.43 & 0.925 $\pm$ 0.020 & 1.063 $\pm$ 0.041 & 0 & 0 &  ok  \\ 
OGLE-LMC-ACEP-004 & F     &   1.86 & 13.12 $\pm$  2.99 & 1.091 $\pm$ 0.202 & 1.202 $\pm$ 0.383 & 0 & 0 &  OK  \\ 
OGLE-LMC-ACEP-005 & F     &   0.93 &  8.53 $\pm$  0.50 & 0.960 $\pm$ 0.017 & 1.106 $\pm$ 0.032 & 0 & 0 &  ok  \\ 
OGLE-LMC-ACEP-006 & 1O    &   1.14 & 12.07 $\pm$  0.71 & 1.703 $\pm$ 0.042 & 2.115 $\pm$ 0.110 & 0 & 0 &      \\ 
OGLE-LMC-ACEP-007 & F     &   0.90 &  8.75 $\pm$  0.66 & 1.096 $\pm$ 0.029 & 1.267 $\pm$ 0.064 & 0 & 0 &  OK  \\ 
OGLE-LMC-ACEP-008 & 1O    &   1.00 & 10.64 $\pm$  0.44 & 1.500 $\pm$ 0.020 & 1.820 $\pm$ 0.043 & 0 & 0 &      \\ 
OGLE-LMC-ACEP-009 & 1O    &   1.07 & 10.56 $\pm$  0.42 & 1.344 $\pm$ 0.016 & 1.605 $\pm$ 0.032 & 0 & 0 &      \\ 
OGLE-LMC-ACEP-010 & F     &   0.83 &  7.69 $\pm$  0.73 & 0.878 $\pm$ 0.029 & 0.993 $\pm$ 0.059 & 0 & 0 &  OK  \\ 
OGLE-LMC-ACEP-011 & F     &   1.00 & 11.44 $\pm$  1.01 & 1.667 $\pm$ 0.085 & 2.253 $\pm$ 0.260 & 0 & 0 &  OK  \\ 
OGLE-LMC-ACEP-012 & F     &   0.83 &  9.99 $\pm$  2.88 & 1.571 $\pm$ 0.619 & 2.083 $\pm$ 1.623 & 0 & 0 &  OK  \\ 
OGLE-LMC-ACEP-013 & 1O    &   0.67 &  7.51 $\pm$  1.08 & 1.090 $\pm$ 0.090 & 1.321 $\pm$ 0.218 & 0 & 0 &  OK  \\ 
OGLE-LMC-ACEP-014 & F     &   2.29 & 15.37 $\pm$  0.63 & 1.233 $\pm$ 0.015 & 1.339 $\pm$ 0.025 & 0 & 0 &  ok  \\ 
OGLE-LMC-ACEP-015 & 1O    &   1.58 & 12.12 $\pm$  0.69 & 1.150 $\pm$ 0.020 & 1.252 $\pm$ 0.038 & 0 & 0 &  OK  \\ 
OGLE-LMC-ACEP-016 & F     &   1.55 & 12.51 $\pm$  0.76 & 1.261 $\pm$ 0.026 & 1.422 $\pm$ 0.053 & 0 & 0 &  OK  \\ 
OGLE-LMC-ACEP-017 & F     &   0.93 &  9.64 $\pm$  0.39 & 1.298 $\pm$ 0.016 & 1.565 $\pm$ 0.031 & 0 & 0 &      \\ 
OGLE-LMC-ACEP-018 & F     &   1.02 &  8.80 $\pm$  0.80 & 0.957 $\pm$ 0.031 & 1.046 $\pm$ 0.061 & 0 & 0 &  OK  \\ 
OGLE-LMC-ACEP-019 & F     &   0.91 &  9.23 $\pm$  0.50 & 1.259 $\pm$ 0.022 & 1.435 $\pm$ 0.045 & 0 & 0 &  ok  \\ 
OGLE-LMC-ACEP-020 & 1O    &   0.51 &  6.14 $\pm$  0.50 & 1.047 $\pm$ 0.031 & 1.161 $\pm$ 0.062 & 0 & 0 &  OK  \\ 
OGLE-LMC-ACEP-021 & F     &   1.30 & 11.22 $\pm$  1.83 & 1.206 $\pm$ 0.136 & 1.396 $\pm$ 0.299 & 0 & 0 &  OK  \\ 
OGLE-LMC-ACEP-023 & 1O    &   0.97 & 11.32 $\pm$  0.69 & 1.811 $\pm$ 0.052 & 2.291 $\pm$ 0.137 & 0 & 0 &      \\ 
OGLE-LMC-ACEP-024 & F     &   0.79 &  6.98 $\pm$  0.24 & 0.825 $\pm$ 0.011 & 0.817 $\pm$ 0.012 & 0 & 0 &  OK  \\ 
OGLE-LMC-ACEP-025 & 1O    &   0.64 &  6.82 $\pm$  0.90 & 0.991 $\pm$ 0.064 & 1.097 $\pm$ 0.129 & 0 & 0 &  OK  \\ 
OGLE-LMC-ACEP-026 & F     &   1.74 & 14.61 $\pm$  0.61 & 1.509 $\pm$ 0.020 & 1.815 $\pm$ 0.044 & 0 & 0 &      \\ 
OGLE-LMC-ACEP-027 & F     &   1.27 & 11.76 $\pm$  1.09 & 1.430 $\pm$ 0.069 & 1.649 $\pm$ 0.154 & 0 & 0 &  OK  \\ 
OGLE-LMC-ACEP-028 & 1O    &   0.80 & 12.17 $\pm$  1.33 & 2.467 $\pm$ 0.276 & 3.821 $\pm$ 1.117 & 0 & 0 &  OK  \\ 
OGLE-LMC-ACEP-029 & F     &   0.80 &  6.69 $\pm$  0.48 & 0.688 $\pm$ 0.014 & 0.715 $\pm$ 0.021 & 0 & 0 &  OK  \\ 
OGLE-LMC-ACEP-030 & 1O    &   0.89 & 10.40 $\pm$  0.78 & 1.662 $\pm$ 0.063 & 2.060 $\pm$ 0.163 & 0 & 0 &  OK  \\ 
OGLE-LMC-ACEP-031 & 1O    &   1.12 & 10.61 $\pm$  1.34 & 1.293 $\pm$ 0.099 & 1.502 $\pm$ 0.223 & 0 & 0 &  OK  \\ 
OGLE-LMC-ACEP-032 & F     &   1.32 & 12.24 $\pm$  0.52 & 1.432 $\pm$ 0.019 & 1.737 $\pm$ 0.041 & 0 & 0 &      \\ 
OGLE-LMC-ACEP-033 & F     &   2.35 & 15.90 $\pm$  0.66 & 1.278 $\pm$ 0.016 & 1.414 $\pm$ 0.027 & 0 & 0 &  ok  \\ 
OGLE-LMC-ACEP-034 & F     &   0.73 &  8.05 $\pm$  1.75 & 1.156 $\pm$ 0.209 & 1.383 $\pm$ 0.469 & 0 & 0 &  OK  \\ 
OGLE-LMC-ACEP-035 & 1O    &   0.60 &  6.64 $\pm$  0.25 & 0.977 $\pm$ 0.012 & 1.129 $\pm$ 0.017 & 0 & 0 &  ok  \\ 
OGLE-LMC-ACEP-036 & F     &   1.26 & 12.09 $\pm$  0.50 & 1.473 $\pm$ 0.019 & 1.807 $\pm$ 0.042 & 0 & 0 &      \\ 
OGLE-LMC-ACEP-037 & F     &   1.26 & 20.42 $\pm$  2.77 & 4.582 $\pm$ 1.417 & 7.854 $\pm$ 6.898 & 0 & 0 &  OK  \\ 
OGLE-LMC-ACEP-038 & F     &   1.34 & 10.51 $\pm$  0.61 & 1.019 $\pm$ 0.017 & 1.107 $\pm$ 0.031 & 0 & 0 &  OK  \\ 
OGLE-LMC-ACEP-039 & F     &   0.99 &  8.81 $\pm$  0.18 & 0.981 $\pm$ 0.010 & 1.095 $\pm$ 0.011 & 0 & 0 &  ok  \\ 
OGLE-LMC-ACEP-040 & F     &   0.96 & 10.81 $\pm$  1.25 & 1.589 $\pm$ 0.128 & 2.044 $\pm$ 0.356 & 0 & 0 &  OK  \\ 
OGLE-LMC-ACEP-041 & F     &   0.88 &  8.57 $\pm$  1.64 & 1.058 $\pm$ 0.141 & 1.235 $\pm$ 0.306 & 0 & 0 &  OK  \\ 
OGLE-LMC-ACEP-042 & F     &   1.08 & 12.10 $\pm$  0.63 & 1.613 $\pm$ 0.032 & 2.329 $\pm$ 0.105 & 0 & 0 &      \\ 
OGLE-LMC-ACEP-043 & 1O    &   0.68 &  8.53 $\pm$  0.35 & 1.414 $\pm$ 0.018 & 1.854 $\pm$ 0.044 & 0 & 0 &      \\ 
OGLE-LMC-ACEP-044 & F     &   1.31 & 11.09 $\pm$  0.99 & 1.224 $\pm$ 0.048 & 1.327 $\pm$ 0.094 & 0 & 0 &  OK  \\ 
OGLE-LMC-ACEP-045 & F     &   0.68 &  6.83 $\pm$  0.53 & 0.851 $\pm$ 0.020 & 0.994 $\pm$ 0.042 & 0 & 0 &  ok  \\ 
OGLE-LMC-ACEP-046 & F     &   1.26 & 11.00 $\pm$  1.58 & 1.214 $\pm$ 0.110 & 1.374 $\pm$ 0.233 & 0 & 0 &  OK  \\ 
OGLE-LMC-ACEP-047 & F     &   2.18 & 12.63 $\pm$  0.74 & 0.849 $\pm$ 0.014 & 0.838 $\pm$ 0.020 & 0 & 0 &  OK  \\ 
OGLE-LMC-ACEP-048 & F     &   1.55 & 14.07 $\pm$  1.88 & 1.622 $\pm$ 0.173 & 1.978 $\pm$ 0.428 & 0 & 0 &  OK  \\ 
OGLE-LMC-ACEP-049 & F     &   0.64 &  6.67 $\pm$  0.99 & 0.915 $\pm$ 0.068 & 1.008 $\pm$ 0.134 & 0 & 0 &  OK  \\ 
OGLE-LMC-ACEP-050 & 1O    &   1.40 & 13.40 $\pm$  0.51 & 1.750 $\pm$ 0.023 & 2.022 $\pm$ 0.046 & 0 & 0 &  ok  \\ 
OGLE-LMC-ACEP-051 & F     &   0.71 &  7.32 $\pm$  0.15 & 0.949 $\pm$ 0.010 & 1.128 $\pm$ 0.011 & 0 & 0 &      \\ 
OGLE-LMC-ACEP-052 & F     &   1.26 & 12.30 $\pm$  1.20 & 1.544 $\pm$ 0.088 & 1.885 $\pm$ 0.222 & 0 & 0 &  OK  \\ 
OGLE-LMC-ACEP-053 & F     &   1.89 & 16.49 $\pm$  1.99 & 1.785 $\pm$ 0.174 & 2.231 $\pm$ 0.456 & 0 & 0 &  OK  \\ 
OGLE-LMC-ACEP-054 & F     &   0.98 & 10.45 $\pm$  0.70 & 1.346 $\pm$ 0.035 & 1.804 $\pm$ 0.103 & 0 & 0 &      \\ 
OGLE-LMC-ACEP-055 & F     &   1.61 & 13.24 $\pm$  1.58 & 1.326 $\pm$ 0.094 & 1.569 $\pm$ 0.220 & 0 & 0 &  OK  \\ 
OGLE-LMC-ACEP-056 & F     &   1.12 & 10.68 $\pm$  1.57 & 1.284 $\pm$ 0.129 & 1.532 $\pm$ 0.304 & 0 & 0 &  OK  \\ 
OGLE-LMC-ACEP-057 & F     &   1.71 & 14.65 $\pm$  0.89 & 1.572 $\pm$ 0.039 & 1.882 $\pm$ 0.094 & 0 & 0 &      \\ 
OGLE-LMC-ACEP-058 & 1O    &   0.65 &  6.40 $\pm$  1.16 & 0.889 $\pm$ 0.090 & 0.885 $\pm$ 0.143 & 0 & 0 &  OK  \\ 
OGLE-LMC-ACEP-059 & F     &   0.83 & 10.67 $\pm$  0.86 & 1.564 $\pm$ 0.064 & 2.490 $\pm$ 0.274 & 0 & 0 &      \\ 
OGLE-LMC-ACEP-060 & F     &   1.28 & 12.21 $\pm$  0.91 & 1.532 $\pm$ 0.053 & 1.814 $\pm$ 0.127 & 0 & 0 &  OK  \\ 
\hline
\end{tabular}
\end{table*}

\setcounter{table}{0}
\begin{table*}
\caption{Continued}

\begin{tabular}{lrrrrrrrrlrrrrr}
\hline
Name               & Type  & Period & Radius   & Mass$_{\rm Cep}$  & Mass$_{\rm RRL}$    & Dusty? & Binary? & Agree? \\ 
                   &       &  (d)   &  (\rsol) &     (\msol)     &     (\msol)       &        &         &            \\
\hline

OGLE-LMC-ACEP-061 & F     &   0.85 &  8.84 $\pm$  0.52 & 1.175 $\pm$ 0.022 & 1.425 $\pm$ 0.052 & 0 & 0 &      \\ 
OGLE-LMC-ACEP-062 & F     &   1.06 & 10.02 $\pm$  1.33 & 1.284 $\pm$ 0.107 & 1.407 $\pm$ 0.213 & 0 & 0 &  OK  \\ 
OGLE-LMC-ACEP-063 & F     &   0.89 &  8.86 $\pm$  0.75 & 1.055 $\pm$ 0.033 & 1.320 $\pm$ 0.085 & 0 & 0 &  OK  \\ 
OGLE-LMC-ACEP-064 & F     &   1.36 & 12.17 $\pm$  0.93 & 1.392 $\pm$ 0.046 & 1.627 $\pm$ 0.106 & 0 & 0 &  OK  \\ 
OGLE-LMC-ACEP-065 & F     &   1.32 & 11.03 $\pm$  1.38 & 1.169 $\pm$ 0.080 & 1.289 $\pm$ 0.161 & 0 & 0 &  OK  \\ 
OGLE-LMC-ACEP-066 & F     &   1.04 &  9.50 $\pm$  0.72 & 1.089 $\pm$ 0.029 & 1.252 $\pm$ 0.063 & 0 & 0 &  OK  \\ 
OGLE-LMC-ACEP-067 & F     &   0.82 & 10.16 $\pm$  3.10 & 1.635 $\pm$ 0.735 & 2.221 $\pm$ 2.000 & 0 & 0 &  OK  \\ 
OGLE-LMC-ACEP-068 & F     &   0.63 &  5.68 $\pm$  0.85 & 0.652 $\pm$ 0.036 & 0.674 $\pm$ 0.061 & 0 & 0 &  OK  \\ 
OGLE-LMC-ACEP-069 & F     &   1.54 & 15.40 $\pm$  1.25 & 1.991 $\pm$ 0.104 & 2.567 $\pm$ 0.295 & 0 & 0 &  OK  \\ 
OGLE-LMC-ACEP-070 & 1O    &   0.84 &  7.82 $\pm$  0.71 & 0.918 $\pm$ 0.029 & 1.019 $\pm$ 0.058 & 0 & 0 &  OK  \\ 
OGLE-LMC-ACEP-071 & 1O    &   0.91 &  8.59 $\pm$  0.88 & 1.102 $\pm$ 0.049 & 1.180 $\pm$ 0.094 & 0 & 0 &  OK  \\ 
OGLE-LMC-ACEP-072 & F     &   1.05 & 11.21 $\pm$  1.21 & 1.622 $\pm$ 0.117 & 1.961 $\pm$ 0.289 & 0 & 0 &  OK  \\ 
OGLE-LMC-ACEP-073 & F     &   1.47 & 13.12 $\pm$  1.65 & 1.523 $\pm$ 0.137 & 1.770 $\pm$ 0.310 & 0 & 0 &  OK  \\ 
OGLE-LMC-ACEP-074 & F     &   1.53 & 12.69 $\pm$  0.74 & 1.317 $\pm$ 0.026 & 1.500 $\pm$ 0.055 & 0 & 0 &  ok  \\ 
OGLE-LMC-ACEP-075 & F     &   0.69 &  7.40 $\pm$  0.67 & 1.034 $\pm$ 0.036 & 1.203 $\pm$ 0.080 & 0 & 0 &  OK  \\ 
OGLE-LMC-ACEP-076 & F     &   1.58 & 11.87 $\pm$  1.11 & 1.083 $\pm$ 0.041 & 1.182 $\pm$ 0.082 & 0 & 0 &  OK  \\ 
OGLE-LMC-ACEP-077 & F     &   1.12 & 10.98 $\pm$  0.50 & 1.342 $\pm$ 0.019 & 1.662 $\pm$ 0.043 & 0 & 0 &      \\ 
OGLE-LMC-ACEP-078 & 1O    &   1.15 & 10.84 $\pm$  0.58 & 1.383 $\pm$ 0.025 & 1.542 $\pm$ 0.050 & 0 & 0 &  OK  \\ 
OGLE-LMC-ACEP-079 & F     &   1.16 & 10.80 $\pm$  0.82 & 1.330 $\pm$ 0.042 & 1.509 $\pm$ 0.089 & 0 & 0 &  OK  \\ 
OGLE-LMC-ACEP-080 & F     &   1.06 &  9.83 $\pm$  0.74 & 1.169 $\pm$ 0.032 & 1.342 $\pm$ 0.070 & 0 & 0 &  OK  \\ 
OGLE-LMC-ACEP-081 & F     &   0.80 &  7.75 $\pm$  0.45 & 0.949 $\pm$ 0.016 & 1.081 $\pm$ 0.030 & 0 & 0 &  ok  \\ 
OGLE-LMC-ACEP-082 & 1O    &   1.04 & 13.07 $\pm$  0.58 & 2.166 $\pm$ 0.041 & 3.070 $\pm$ 0.137 & 0 & 0 &      \\ 
OGLE-LMC-T2CEP-001 & BLHer &   1.81 &  9.31 $\pm$  0.20 & 0.507 $\pm$ 0.010 & 0.481 $\pm$ 0.010 & 0 & 0 &  OK  \\ 
OGLE-LMC-T2CEP-002 & WVir  &  18.32 & 30.37 $\pm$  3.39 & 0.411 $\pm$ 0.013 & 0.310 $\pm$ 0.013 & 0 & 0 &      \\ 
OGLE-LMC-T2CEP-003 & RVTau &  35.66 & 60.28 $\pm$  8.03 & 0.953 $\pm$ 0.060 & 0.714 $\pm$ 0.056 & 1 & 0 &  OK  \\ 
OGLE-LMC-T2CEP-004 & BLHer &   1.92 & 12.62 $\pm$  1.84 & 0.928 $\pm$ 0.066 & 1.032 $\pm$ 0.134 & 0 & 0 &  OK  \\ 
OGLE-LMC-T2CEP-005 & RVTau &  33.19 & 50.18 $\pm$  6.41 & 0.605 $\pm$ 0.024 & 0.483 $\pm$ 0.025 & 0 & 0 &      \\ 
OGLE-LMC-T2CEP-006 & BLHer &   1.09 &  6.26 $\pm$  0.54 & 0.414 $\pm$ 0.011 & 0.361 $\pm$ 0.012 & 0 & 0 &  ok  \\ 
OGLE-LMC-T2CEP-007 & BLHer &   1.24 &  6.82 $\pm$  0.49 & 0.421 $\pm$ 0.011 & 0.372 $\pm$ 0.011 & 0 & 0 &  ok  \\ 
OGLE-LMC-T2CEP-008 & BLHer &   1.75 &  9.19 $\pm$  0.39 & 0.509 $\pm$ 0.010 & 0.493 $\pm$ 0.011 & 0 & 0 &  OK  \\ 
OGLE-LMC-T2CEP-009 & BLHer &   1.76 &  8.71 $\pm$  0.35 & 0.460 $\pm$ 0.010 & 0.418 $\pm$ 0.010 & 0 & 0 &  OK  \\ 
OGLE-LMC-T2CEP-010 & BLHer &   1.50 &  7.35 $\pm$  0.54 & 0.386 $\pm$ 0.011 & 0.336 $\pm$ 0.011 & 0 & 0 &  ok  \\ 
OGLE-LMC-T2CEP-011 & RVTau &  39.26 & 52.01 $\pm$  2.21 & 0.592 $\pm$ 0.010 & 0.405 $\pm$ 0.010 & 1 & 0 &      \\ 
OGLE-LMC-T2CEP-012 & WVir  &  11.58 & 24.80 $\pm$  2.23 & 0.454 $\pm$ 0.012 & 0.370 $\pm$ 0.012 & 0 & 0 &      \\ 
OGLE-LMC-T2CEP-013 & WVir  &  11.54 & 23.84 $\pm$  1.12 & 0.416 $\pm$ 0.010 & 0.333 $\pm$ 0.010 & 0 & 0 &      \\ 
OGLE-LMC-T2CEP-014 & RVTau &  61.88 & 48.67 $\pm$  2.12 & 0.283 $\pm$ 0.010 & 0.161 $\pm$ 0.010 & 1 & 0 &      \\ 
OGLE-LMC-T2CEP-015 & RVTau &  56.52 & 72.01 $\pm$  3.53 & 0.732 $\pm$ 0.011 & 0.559 $\pm$ 0.011 & 1 & 0 &      \\ 
OGLE-LMC-T2CEP-016 & RVTau &  20.30 & 23.45 $\pm$  2.12 & 0.228 $\pm$ 0.010 & 0.127 $\pm$ 0.010 & 1 & 0 &      \\ 
OGLE-LMC-T2CEP-017 & WVir  &  14.45 & 30.64 $\pm$  2.31 & 0.541 $\pm$ 0.012 & 0.467 $\pm$ 0.013 & 0 & 0 &      \\ 
OGLE-LMC-T2CEP-018 & BLHer &   1.38 &  7.70 $\pm$  0.30 & 0.474 $\pm$ 0.010 & 0.440 $\pm$ 0.010 & 0 & 0 &  OK  \\ 
OGLE-LMC-T2CEP-019 & pWVir &   8.67 & 27.86 $\pm$  6.88 & 0.833 $\pm$ 0.135 & 0.819 $\pm$ 0.201 & 0 & 0 &  OK  \\ 
OGLE-LMC-T2CEP-020 & BLHer &   1.11 &  7.60 $\pm$  0.70 & 0.611 $\pm$ 0.016 & 0.605 $\pm$ 0.023 & 0 & 0 &  OK  \\ 
OGLE-LMC-T2CEP-021 & pWVir &   9.76 & 23.72 $\pm$  1.04 & 0.534 $\pm$ 0.010 & 0.430 $\pm$ 0.010 & 0 & 1 &      \\ 
OGLE-LMC-T2CEP-022 & WVir  &  10.72 & 23.70 $\pm$  1.62 & 0.450 $\pm$ 0.011 & 0.369 $\pm$ 0.011 & 0 & 0 &      \\ 
OGLE-LMC-T2CEP-023 & pWVir &   5.23 & 24.72 $\pm$  1.90 & 1.343 $\pm$ 0.044 & 1.323 $\pm$ 0.072 & 0 & 1 &  OK  \\ 
OGLE-LMC-T2CEP-024 & BLHer &   1.25 &  6.85 $\pm$  0.74 & 0.413 $\pm$ 0.012 & 0.373 $\pm$ 0.014 & 0 & 0 &  OK  \\ 
OGLE-LMC-T2CEP-025 & RVTau &  67.97 & 75.77 $\pm$  9.15 & 0.645 $\pm$ 0.025 & 0.478 $\pm$ 0.023 & 0 & 0 &      \\ 
OGLE-LMC-T2CEP-026 & WVir  &  13.58 & 28.09 $\pm$  1.39 & 0.485 $\pm$ 0.010 & 0.406 $\pm$ 0.010 & 0 & 0 &      \\ 
OGLE-LMC-T2CEP-027 & WVir  &  17.13 & 27.46 $\pm$  3.44 & 0.362 $\pm$ 0.013 & 0.260 $\pm$ 0.012 & 0 & 0 &      \\ 
OGLE-LMC-T2CEP-028 & pWVir &   8.78 & 23.71 $\pm$  2.61 & 0.645 $\pm$ 0.022 & 0.508 $\pm$ 0.022 & 0 & 0 &      \\ 
OGLE-LMC-T2CEP-029 & RVTau &  31.25 & 53.89 $\pm$  3.43 & 0.845 $\pm$ 0.015 & 0.648 $\pm$ 0.016 & 1 & 0 &      \\ 
OGLE-LMC-T2CEP-030 & BLHer &   3.94 & 14.34 $\pm$  0.63 & 0.514 $\pm$ 0.010 & 0.459 $\pm$ 0.010 & 0 & 0 &  ok  \\ 
OGLE-LMC-T2CEP-031 & WVir  &   6.71 & 18.38 $\pm$  0.84 & 0.454 $\pm$ 0.010 & 0.388 $\pm$ 0.010 & 0 & 0 &      \\ 
OGLE-LMC-T2CEP-032 & RVTau &  44.56 & 96.44 $\pm$ 32.06 & 1.865 $\pm$ 1.092 & 1.863 $\pm$ 1.571 & 1 & 0 &  OK  \\ 
OGLE-LMC-T2CEP-033 & pWVir &   9.39 & 23.79 $\pm$  1.92 & 0.571 $\pm$ 0.013 & 0.461 $\pm$ 0.014 & 0 & 0 &      \\ 
OGLE-LMC-T2CEP-034 & WVir  &  14.91 & 29.99 $\pm$  1.60 & 0.488 $\pm$ 0.010 & 0.419 $\pm$ 0.011 & 0 & 0 &      \\ 
OGLE-LMC-T2CEP-035 & WVir  &   9.87 & 26.15 $\pm$  2.49 & 0.612 $\pm$ 0.017 & 0.557 $\pm$ 0.021 & 0 & 0 &  OK  \\ 
\hline
\end{tabular}
\end{table*}

\setcounter{table}{0}
\begin{table*}
\caption{Continued}

\begin{tabular}{lrrrrrrrrlrrrrr}
\hline
Name               & Type  & Period & Radius   & Mass$_{\rm Cep}$  & Mass$_{\rm RRL}$    & Dusty? & Binary? & Agree? \\ 
                   &       &  (d)   &  (\rsol) &     (\msol)     &     (\msol)       &        &         &            \\
\hline

OGLE-LMC-T2CEP-036 & WVir  &  14.88 & 23.60 $\pm$  1.55 & 0.308 $\pm$ 0.010 & 0.214 $\pm$ 0.010 & 0 & 0 &      \\ 
OGLE-LMC-T2CEP-037 & WVir  &   6.90 & 17.20 $\pm$  0.75 & 0.385 $\pm$ 0.010 & 0.308 $\pm$ 0.010 & 0 & 0 &      \\ 
OGLE-LMC-T2CEP-038 & WVir  &   4.01 & 14.72 $\pm$  1.46 & 0.607 $\pm$ 0.017 & 0.476 $\pm$ 0.018 & 0 & 0 &      \\ 
OGLE-LMC-T2CEP-039 & WVir  &   8.72 & 20.20 $\pm$  1.72 & 0.418 $\pm$ 0.011 & 0.330 $\pm$ 0.011 & 0 & 0 &      \\ 
OGLE-LMC-T2CEP-040 & pWVir &   9.63 & 33.74 $\pm$  4.61 & 1.143 $\pm$ 0.090 & 1.182 $\pm$ 0.159 & 0 & 0 &  OK  \\ 
OGLE-LMC-T2CEP-041 & BLHer &   2.48 & 10.81 $\pm$  1.53 & 0.541 $\pm$ 0.023 & 0.440 $\pm$ 0.025 & 0 & 0 &  OK  \\ 
OGLE-LMC-T2CEP-042 & pWVir &   4.92 & 14.36 $\pm$  2.17 & 0.427 $\pm$ 0.018 & 0.320 $\pm$ 0.017 & 0 & 0 &      \\ 
OGLE-LMC-T2CEP-043 & WVir  &   6.56 & 16.79 $\pm$  1.13 & 0.378 $\pm$ 0.010 & 0.312 $\pm$ 0.010 & 0 & 0 &      \\ 
OGLE-LMC-T2CEP-044 & WVir  &  13.27 & 25.38 $\pm$  1.74 & 0.405 $\pm$ 0.010 & 0.316 $\pm$ 0.011 & 0 & 0 &      \\ 
OGLE-LMC-T2CEP-045 & RVTau &  63.39 & 74.94 $\pm$  3.79 & 0.706 $\pm$ 0.011 & 0.519 $\pm$ 0.011 & 0 & 0 &      \\ 
OGLE-LMC-T2CEP-046 & WVir  &  14.74 & 35.90 $\pm$  1.70 & 0.797 $\pm$ 0.012 & 0.704 $\pm$ 0.013 & 1 & 0 &  ok  \\ 
OGLE-LMC-T2CEP-047 & WVir  &   7.29 & 18.63 $\pm$  1.60 & 0.427 $\pm$ 0.011 & 0.352 $\pm$ 0.012 & 0 & 0 &      \\ 
OGLE-LMC-T2CEP-048 & BLHer &   1.45 &  7.89 $\pm$  1.71 & 0.473 $\pm$ 0.036 & 0.436 $\pm$ 0.048 & 0 & 0 &  OK  \\ 
OGLE-LMC-T2CEP-049 & BLHer &   3.24 & 12.47 $\pm$  2.71 & 0.503 $\pm$ 0.041 & 0.426 $\pm$ 0.045 & 0 & 0 &  OK  \\ 
OGLE-LMC-T2CEP-050 & RVTau &  34.75 & 36.52 $\pm$  1.57 & 0.302 $\pm$ 0.010 & 0.184 $\pm$ 0.010 & 1 & 0 &      \\ 
OGLE-LMC-T2CEP-051 & RVTau &  40.61 & 47.45 $\pm$  3.14 & 0.441 $\pm$ 0.011 & 0.297 $\pm$ 0.010 & 0 & 0 &      \\ 
OGLE-LMC-T2CEP-052 & pWVir &   4.69 & 14.41 $\pm$  1.91 & 0.466 $\pm$ 0.017 & 0.349 $\pm$ 0.016 & 0 & 1 &      \\ 
OGLE-LMC-T2CEP-053 & BLHer &   1.04 &  6.84 $\pm$  0.38 & 0.520 $\pm$ 0.011 & 0.496 $\pm$ 0.011 & 0 & 0 &  OK  \\ 
OGLE-LMC-T2CEP-054 & WVir  &   9.93 & 23.35 $\pm$  0.57 & 0.473 $\pm$ 0.010 & 0.401 $\pm$ 0.010 & 0 & 0 &      \\ 
OGLE-LMC-T2CEP-055 & RVTau &  41.01 & 50.93 $\pm$  3.26 & 0.527 $\pm$ 0.011 & 0.356 $\pm$ 0.011 & 1 & 0 &      \\ 
OGLE-LMC-T2CEP-056 & WVir  &   7.29 & 19.92 $\pm$  0.49 & 0.480 $\pm$ 0.010 & 0.425 $\pm$ 0.010 & 0 & 0 &  ok  \\ 
OGLE-LMC-T2CEP-057 & WVir  &  16.63 & 30.31 $\pm$  3.42 & 0.456 $\pm$ 0.014 & 0.361 $\pm$ 0.015 & 0 & 0 &      \\ 
OGLE-LMC-T2CEP-058 & RVTau &  21.48 & 33.97 $\pm$  3.83 & 0.431 $\pm$ 0.013 & 0.328 $\pm$ 0.013 & 0 & 0 &      \\ 
OGLE-LMC-T2CEP-059 & WVir  &  16.74 & 34.09 $\pm$  3.16 & 0.595 $\pm$ 0.015 & 0.496 $\pm$ 0.017 & 0 & 0 &      \\ 
OGLE-LMC-T2CEP-060 & BLHer &   1.24 &  7.30 $\pm$  1.39 & 0.474 $\pm$ 0.030 & 0.453 $\pm$ 0.042 & 0 & 0 &  OK  \\ 
OGLE-LMC-T2CEP-061 & BLHer &   1.18 &  5.96 $\pm$  0.86 & 0.336 $\pm$ 0.013 & 0.275 $\pm$ 0.014 & 0 & 0 &      \\ 
OGLE-LMC-T2CEP-062 & WVir  &   6.05 & 20.36 $\pm$  3.43 & 0.612 $\pm$ 0.039 & 0.612 $\pm$ 0.061 & 0 & 0 &  OK  \\ 
OGLE-LMC-T2CEP-063 & WVir  &   6.92 & 17.59 $\pm$  1.49 & 0.403 $\pm$ 0.011 & 0.325 $\pm$ 0.011 & 0 & 0 &      \\ 
OGLE-LMC-T2CEP-064 & BLHer &   2.13 &  9.80 $\pm$  2.05 & 0.473 $\pm$ 0.034 & 0.429 $\pm$ 0.044 & 0 & 0 &  OK  \\ 
OGLE-LMC-T2CEP-065 & RVTau &  35.05 & 45.67 $\pm$  2.07 & 0.479 $\pm$ 0.010 & 0.339 $\pm$ 0.010 & 1 & 0 &      \\ 
OGLE-LMC-T2CEP-066 & WVir  &  13.11 & 25.80 $\pm$  0.64 & 0.421 $\pm$ 0.010 & 0.338 $\pm$ 0.010 & 0 & 0 &      \\ 
OGLE-LMC-T2CEP-067 & RVTau &  48.23 & 71.33 $\pm$ 10.49 & 0.978 $\pm$ 0.075 & 0.701 $\pm$ 0.064 & 1 & 0 &  OK  \\ 
OGLE-LMC-T2CEP-068 & BLHer &   1.61 &  8.14 $\pm$  0.60 & 0.449 $\pm$ 0.011 & 0.400 $\pm$ 0.012 & 0 & 0 &  ok  \\ 
OGLE-LMC-T2CEP-069 & BLHer &   1.02 &  7.34 $\pm$  1.96 & 0.630 $\pm$ 0.088 & 0.626 $\pm$ 0.130 & 0 & 0 &  OK  \\ 
OGLE-LMC-T2CEP-070 & WVir  &  15.44 & 25.12 $\pm$  2.56 & 0.348 $\pm$ 0.011 & 0.240 $\pm$ 0.011 & 0 & 0 &      \\ 
OGLE-LMC-T2CEP-071 & BLHer &   1.15 &  7.14 $\pm$  1.67 & 0.497 $\pm$ 0.045 & 0.476 $\pm$ 0.063 & 0 & 0 &  OK  \\ 
OGLE-LMC-T2CEP-072 & WVir  &  14.51 & 26.70 $\pm$  1.23 & 0.412 $\pm$ 0.010 & 0.315 $\pm$ 0.010 & 0 & 0 &      \\ 
OGLE-LMC-T2CEP-073 & BLHer &   3.09 & 12.58 $\pm$  1.02 & 0.519 $\pm$ 0.012 & 0.471 $\pm$ 0.014 & 0 & 0 &  OK  \\ 
OGLE-LMC-T2CEP-074 & WVir  &   8.99 & 22.53 $\pm$  1.45 & 0.519 $\pm$ 0.011 & 0.426 $\pm$ 0.011 & 0 & 0 &      \\ 
OGLE-LMC-T2CEP-075 & RVTau &  50.19 & 54.91 $\pm$  2.77 & 0.458 $\pm$ 0.010 & 0.317 $\pm$ 0.010 & 1 & 0 &      \\ 
OGLE-LMC-T2CEP-076 & BLHer &   2.10 & 10.40 $\pm$  1.10 & 0.519 $\pm$ 0.015 & 0.516 $\pm$ 0.022 & 0 & 0 &  OK  \\ 
OGLE-LMC-T2CEP-077 & BLHer &   1.21 &  6.82 $\pm$  0.65 & 0.460 $\pm$ 0.012 & 0.385 $\pm$ 0.013 & 0 & 1 &      \\ 
OGLE-LMC-T2CEP-078 & pWVir &   6.72 & 28.22 $\pm$  1.48 & 1.165 $\pm$ 0.019 & 1.285 $\pm$ 0.034 & 0 & 0 &  ok  \\ 
OGLE-LMC-T2CEP-079 & WVir  &  14.85 & 26.06 $\pm$  2.02 & 0.359 $\pm$ 0.010 & 0.285 $\pm$ 0.011 & 0 & 0 &      \\ 
OGLE-LMC-T2CEP-080 & RVTau &  40.92 & 51.62 $\pm$  4.44 & 0.539 $\pm$ 0.013 & 0.371 $\pm$ 0.012 & 1 & 0 &      \\ 
OGLE-LMC-T2CEP-081 & WVir  &   9.48 & 22.20 $\pm$  1.02 & 0.457 $\pm$ 0.010 & 0.375 $\pm$ 0.010 & 0 & 0 &      \\ 
OGLE-LMC-T2CEP-082 & RVTau &  35.12 & 42.65 $\pm$  4.96 & 0.397 $\pm$ 0.013 & 0.279 $\pm$ 0.012 & 0 & 0 &      \\ 
OGLE-LMC-T2CEP-083 & pWVir &   5.97 & 17.78 $\pm$  0.79 & 0.498 $\pm$ 0.010 & 0.427 $\pm$ 0.010 & 0 & 0 &      \\ 
OGLE-LMC-T2CEP-084 & BLHer &   1.77 &  8.96 $\pm$  2.84 & 0.551 $\pm$ 0.089 & 0.447 $\pm$ 0.085 & 0 & 1 &  OK  \\ 
OGLE-LMC-T2CEP-085 & BLHer &   3.41 & 11.36 $\pm$  1.27 & 0.375 $\pm$ 0.012 & 0.302 $\pm$ 0.012 & 0 & 0 &      \\ 
OGLE-LMC-T2CEP-086 & WVir  &  15.85 & 28.45 $\pm$  1.87 & 0.434 $\pm$ 0.011 & 0.326 $\pm$ 0.011 & 0 & 0 &      \\ 
OGLE-LMC-T2CEP-087 & WVir  &   5.18 & 16.11 $\pm$  2.31 & 0.466 $\pm$ 0.019 & 0.407 $\pm$ 0.023 & 0 & 0 &  OK  \\ 
OGLE-LMC-T2CEP-088 & BLHer &   1.95 &  7.79 $\pm$  0.71 & 0.358 $\pm$ 0.011 & 0.258 $\pm$ 0.011 & 0 & 0 &      \\ 
OGLE-LMC-T2CEP-089 & BLHer &   1.17 &  6.87 $\pm$  0.37 & 0.462 $\pm$ 0.010 & 0.419 $\pm$ 0.011 & 0 & 0 &  OK  \\ 
OGLE-LMC-T2CEP-090 & BLHer &   1.48 &  8.38 $\pm$  0.64 & 0.524 $\pm$ 0.012 & 0.499 $\pm$ 0.014 & 0 & 0 &  OK  \\ 
\hline
\end{tabular}
\end{table*}

\setcounter{table}{0}
\begin{table*}
\caption{Continued}

\begin{tabular}{lrrrrrrrrlrrrrr}
\hline
Name               & Type  & Period & Radius   & Mass$_{\rm Cep}$  & Mass$_{\rm RRL}$    & Dusty? & Binary? & Agree? \\ 
                   &       &  (d)   &  (\rsol) &     (\msol)     &     (\msol)       &        &         &            \\
\hline

OGLE-LMC-T2CEP-091 & RVTau &  35.75 & 47.36 $\pm$  8.04 & 0.572 $\pm$ 0.034 & 0.362 $\pm$ 0.024 & 1 & 0 &      \\ 
OGLE-LMC-T2CEP-092 & BLHer &   2.62 & 10.68 $\pm$  1.94 & 0.441 $\pm$ 0.024 & 0.390 $\pm$ 0.030 & 0 & 0 &  OK  \\ 
OGLE-LMC-T2CEP-093 & WVir  &  17.59 & 33.65 $\pm$  2.77 & 0.585 $\pm$ 0.014 & 0.440 $\pm$ 0.013 & 0 & 1 &      \\ 
OGLE-LMC-T2CEP-094 & WVir  &   8.47 & 22.54 $\pm$  1.13 & 0.524 $\pm$ 0.010 & 0.471 $\pm$ 0.011 & 0 & 0 &  ok  \\ 
OGLE-LMC-T2CEP-095 & WVir  &   5.00 & 15.79 $\pm$  0.38 & 0.460 $\pm$ 0.010 & 0.408 $\pm$ 0.010 & 0 & 0 &  ok  \\ 
OGLE-LMC-T2CEP-096 & WVir  &  13.93 & 25.78 $\pm$  2.78 & 0.400 $\pm$ 0.012 & 0.306 $\pm$ 0.012 & 0 & 0 &      \\ 
OGLE-LMC-T2CEP-097 & WVir  &  10.51 & 22.69 $\pm$  1.95 & 0.428 $\pm$ 0.011 & 0.338 $\pm$ 0.011 & 0 & 0 &      \\ 
OGLE-LMC-T2CEP-098 & pWVir &   4.97 & 32.80 $\pm$  2.78 & 3.038 $\pm$ 0.263 & 3.159 $\pm$ 0.478 & 0 & 1 &  OK  \\ 
OGLE-LMC-T2CEP-099 & WVir  &  15.49 & 35.43 $\pm$  2.77 & 0.677 $\pm$ 0.015 & 0.628 $\pm$ 0.019 & 0 & 0 &  OK  \\ 
OGLE-LMC-T2CEP-100 & WVir  &   7.43 & 15.73 $\pm$  0.97 & 0.292 $\pm$ 0.010 & 0.212 $\pm$ 0.010 & 0 & 0 &      \\ 
OGLE-LMC-T2CEP-101 & WVir  &  11.42 & 21.60 $\pm$  2.54 & 0.357 $\pm$ 0.012 & 0.257 $\pm$ 0.012 & 0 & 0 &      \\ 
OGLE-LMC-T2CEP-102 & BLHer &   1.27 &  7.57 $\pm$  0.90 & 0.528 $\pm$ 0.018 & 0.481 $\pm$ 0.023 & 0 & 0 &  OK  \\ 
OGLE-LMC-T2CEP-103 & WVir  &  12.91 & 24.61 $\pm$  2.16 & 0.395 $\pm$ 0.011 & 0.304 $\pm$ 0.011 & 0 & 0 &      \\ 
OGLE-LMC-T2CEP-104 & RVTau &  24.88 & 47.95 $\pm$  6.84 & 0.835 $\pm$ 0.052 & 0.676 $\pm$ 0.057 & 1 & 0 &  OK  \\ 
OGLE-LMC-T2CEP-105 & BLHer &   1.49 &  8.54 $\pm$  1.45 & 0.555 $\pm$ 0.033 & 0.520 $\pm$ 0.046 & 0 & 0 &  OK  \\ 
OGLE-LMC-T2CEP-106 & WVir  &   6.71 & 18.21 $\pm$  1.20 & 0.450 $\pm$ 0.011 & 0.378 $\pm$ 0.011 & 0 & 0 &      \\ 
OGLE-LMC-T2CEP-107 & BLHer &   1.21 &  9.91 $\pm$  1.67 & 0.962 $\pm$ 0.093 & 1.106 $\pm$ 0.200 & 0 & 0 &  OK  \\ 
OGLE-LMC-T2CEP-108 & RVTau &  30.01 & 41.06 $\pm$  1.79 & 0.471 $\pm$ 0.010 & 0.323 $\pm$ 0.010 & 0 & 0 &      \\ 
OGLE-LMC-T2CEP-109 & BLHer &   1.41 &  8.23 $\pm$  0.50 & 0.420 $\pm$ 0.010 & 0.513 $\pm$ 0.012 & 0 & 0 &      \\ 
OGLE-LMC-T2CEP-110 & WVir  &   7.08 & 18.84 $\pm$  1.70 & 0.443 $\pm$ 0.012 & 0.381 $\pm$ 0.013 & 0 & 0 &      \\ 
OGLE-LMC-T2CEP-111 & WVir  &   7.50 & 18.76 $\pm$  0.84 & 0.419 $\pm$ 0.010 & 0.343 $\pm$ 0.010 & 0 & 0 &      \\ 
OGLE-LMC-T2CEP-112 & RVTau &  39.40 & 52.33 $\pm$  3.32 & 0.605 $\pm$ 0.012 & 0.409 $\pm$ 0.011 & 1 & 0 &      \\ 
OGLE-LMC-T2CEP-113 & BLHer &   3.09 & 12.43 $\pm$  2.30 & 0.540 $\pm$ 0.036 & 0.455 $\pm$ 0.040 & 0 & 0 &  OK  \\ 
OGLE-LMC-T2CEP-114 & F     &   1.09 & 10.01 $\pm$  0.69 & 1.051 $\pm$ 0.023 & 1.345 $\pm$ 0.060 & 0 & 0 &      \\ 
OGLE-LMC-T2CEP-115 & RVTau &  24.97 & 37.00 $\pm$  2.71 & 0.430 $\pm$ 0.011 & 0.326 $\pm$ 0.011 & 0 & 0 &      \\ 
OGLE-LMC-T2CEP-116 & BLHer &   1.97 &  9.80 $\pm$  1.04 & 0.491 $\pm$ 0.014 & 0.488 $\pm$ 0.020 & 0 & 0 &  OK  \\ 
OGLE-LMC-T2CEP-117 & WVir  &   6.63 & 17.74 $\pm$  0.80 & 0.429 $\pm$ 0.010 & 0.358 $\pm$ 0.010 & 0 & 0 &      \\ 
OGLE-LMC-T2CEP-118 & WVir  &  12.70 & 26.27 $\pm$  1.86 & 0.457 $\pm$ 0.011 & 0.375 $\pm$ 0.011 & 0 & 0 &      \\ 
OGLE-LMC-T2CEP-119 & RVTau &  33.83 & 49.26 $\pm$  8.80 & 0.650 $\pm$ 0.048 & 0.442 $\pm$ 0.036 & 1 & 0 &      \\ 
OGLE-LMC-T2CEP-120 & WVir  &   4.56 & 15.01 $\pm$  0.99 & 0.464 $\pm$ 0.011 & 0.412 $\pm$ 0.011 & 0 & 0 &  ok  \\ 
OGLE-LMC-T2CEP-121 & BLHer &   2.06 &  9.87 $\pm$  1.67 & 0.489 $\pm$ 0.026 & 0.461 $\pm$ 0.036 & 0 & 0 &  OK  \\ 
OGLE-LMC-T2CEP-122 & BLHer &   1.54 &  8.00 $\pm$  0.81 & 0.427 $\pm$ 0.012 & 0.411 $\pm$ 0.015 & 0 & 0 &  OK  \\ 
OGLE-LMC-T2CEP-123 & BLHer &   1.00 & 12.26 $\pm$  1.97 & 1.824 $\pm$ 0.305 & 2.722 $\pm$ 1.103 & 0 & 0 &  OK  \\ 
OGLE-LMC-T2CEP-124 & BLHer &   1.73 &  8.44 $\pm$  1.12 & 0.426 $\pm$ 0.015 & 0.393 $\pm$ 0.020 & 0 & 0 &  OK  \\ 
OGLE-LMC-T2CEP-125 & RVTau &  33.03 & 44.16 $\pm$  4.13 & 0.464 $\pm$ 0.012 & 0.340 $\pm$ 0.012 & 0 & 0 &      \\ 
OGLE-LMC-T2CEP-126 & WVir  &  16.33 & 31.95 $\pm$  1.84 & 0.505 $\pm$ 0.011 & 0.432 $\pm$ 0.011 & 1 & 0 &      \\ 
OGLE-LMC-T2CEP-127 & WVir  &  12.67 & 25.53 $\pm$  4.32 & 0.446 $\pm$ 0.022 & 0.347 $\pm$ 0.022 & 1 & 0 &      \\ 
OGLE-LMC-T2CEP-128 & WVir  &  18.49 & 36.69 $\pm$  3.43 & 0.623 $\pm$ 0.017 & 0.519 $\pm$ 0.018 & 0 & 0 &      \\ 
OGLE-LMC-T2CEP-129 & RVTau &  62.51 & 51.88 $\pm$  2.20 & 0.333 $\pm$ 0.010 & 0.189 $\pm$ 0.010 & 1 & 0 &      \\ 
OGLE-LMC-T2CEP-130 & BLHer &   1.94 &  9.09 $\pm$  1.71 & 0.455 $\pm$ 0.027 & 0.402 $\pm$ 0.033 & 0 & 0 &  OK  \\ 
OGLE-LMC-T2CEP-131 & BLHer &   1.41 &  7.49 $\pm$  0.31 & 0.416 $\pm$ 0.010 & 0.391 $\pm$ 0.010 & 0 & 0 &  OK  \\ 
OGLE-LMC-T2CEP-132 & pWVir &  10.02 & 24.52 $\pm$  2.54 & 0.552 $\pm$ 0.016 & 0.453 $\pm$ 0.017 & 0 & 0 &      \\ 
OGLE-LMC-T2CEP-133 & WVir  &   6.28 & 16.55 $\pm$  1.37 & 0.400 $\pm$ 0.011 & 0.321 $\pm$ 0.011 & 0 & 0 &      \\ 
OGLE-LMC-T2CEP-134 & pWVir &   4.08 & 17.91 $\pm$  1.39 & 0.857 $\pm$ 0.020 & 0.807 $\pm$ 0.029 & 0 & 0 &  OK  \\ 
OGLE-LMC-T2CEP-135 & RVTau &  26.52 & 43.20 $\pm$  3.11 & 0.573 $\pm$ 0.012 & 0.457 $\pm$ 0.012 & 0 & 0 &      \\ 
OGLE-LMC-T2CEP-136 & BLHer &   1.32 & 13.48 $\pm$  4.37 & 1.719 $\pm$ 0.895 & 2.260 $\pm$ 2.246 & 0 & 0 &  OK  \\ 
OGLE-LMC-T2CEP-137 & WVir  &   6.36 & 17.18 $\pm$  2.10 & 0.425 $\pm$ 0.014 & 0.350 $\pm$ 0.015 & 0 & 0 &      \\ 
OGLE-LMC-T2CEP-138 & BLHer &   1.39 & 10.25 $\pm$  3.71 & 0.830 $\pm$ 0.247 & 0.968 $\pm$ 0.475 & 0 & 0 &  OK  \\ 
OGLE-LMC-T2CEP-139 & WVir  &  14.78 & 29.36 $\pm$  1.46 & 0.483 $\pm$ 0.010 & 0.400 $\pm$ 0.010 & 0 & 0 &      \\ 
OGLE-LMC-T2CEP-140 & BLHer &   1.84 &  9.43 $\pm$  0.92 & 0.512 $\pm$ 0.014 & 0.486 $\pm$ 0.018 & 0 & 0 &  OK  \\ 
OGLE-LMC-T2CEP-141 & BLHer &   1.82 &  8.24 $\pm$  1.12 & 0.374 $\pm$ 0.014 & 0.339 $\pm$ 0.016 & 0 & 0 &  OK  \\ 
OGLE-LMC-T2CEP-142 & BLHer &   1.76 & 11.48 $\pm$  0.27 & 0.817 $\pm$ 0.010 & 0.909 $\pm$ 0.011 & 0 & 0 &  ok  \\ 
OGLE-LMC-T2CEP-143 & WVir  &  14.57 & 23.64 $\pm$  2.40 & 0.321 $\pm$ 0.011 & 0.223 $\pm$ 0.011 & 0 & 0 &      \\ 
OGLE-LMC-T2CEP-144 & BLHer &   1.94 & 11.73 $\pm$  4.38 & 0.752 $\pm$ 0.209 & 0.826 $\pm$ 0.352 & 0 & 0 &  OK  \\ 
OGLE-LMC-T2CEP-145 & BLHer &   3.34 & 12.91 $\pm$  2.40 & 0.530 $\pm$ 0.035 & 0.446 $\pm$ 0.039 & 0 & 0 &  OK  \\ 
OGLE-LMC-T2CEP-146 & WVir  &  10.08 & 23.03 $\pm$  1.69 & 0.443 $\pm$ 0.011 & 0.377 $\pm$ 0.011 & 0 & 0 &      \\ 
OGLE-LMC-T2CEP-147 & RVTau &  46.80 & 69.48 $\pm$  6.46 & 0.977 $\pm$ 0.034 & 0.684 $\pm$ 0.029 & 1 & 0 &      \\ 
OGLE-LMC-T2CEP-148 & BLHer &   2.67 &  9.92 $\pm$  0.76 & 0.370 $\pm$ 0.011 & 0.306 $\pm$ 0.011 & 0 & 0 &      \\ 
OGLE-LMC-T2CEP-149 & RVTau &  42.48 & 52.85 $\pm$  4.45 & 0.550 $\pm$ 0.013 & 0.373 $\pm$ 0.012 & 0 & 0 &      \\ 
\hline
\end{tabular}
\end{table*}

\setcounter{table}{0}
\begin{table*}
\caption{Continued}

\begin{tabular}{lrrrrrrrrlrrrrr}
\hline
Name               & Type  & Period & Radius   & Mass$_{\rm Cep}$  & Mass$_{\rm RRL}$    & Dusty? & Binary? & Agree? \\ 
                   &       &  (d)   &  (\rsol) &     (\msol)     &     (\msol)       &        &         &            \\
\hline

OGLE-LMC-T2CEP-150 & WVir  &   5.49 & 17.60 $\pm$  0.67 & 0.585 $\pm$ 0.010 & 0.473 $\pm$ 0.010 & 1 & 0 &      \\ 
OGLE-LMC-T2CEP-151 & WVir  &   7.89 & 19.46 $\pm$  1.67 & 0.428 $\pm$ 0.011 & 0.350 $\pm$ 0.012 & 0 & 0 &      \\ 
OGLE-LMC-T2CEP-152 & WVir  &   9.31 & 22.84 $\pm$  2.52 & 0.493 $\pm$ 0.015 & 0.418 $\pm$ 0.017 & 0 & 0 &      \\ 
OGLE-LMC-T2CEP-153 & BLHer &   1.18 & 11.24 $\pm$  0.71 & 1.591 $\pm$ 0.042 & 1.638 $\pm$ 0.075 & 0 & 0 &  OK  \\ 
OGLE-LMC-T2CEP-154 & pWVir &   7.58 & 23.97 $\pm$  0.92 & 0.822 $\pm$ 0.011 & 0.666 $\pm$ 0.011 & 0 & 0 &      \\ 
OGLE-LMC-T2CEP-155 & WVir  &   6.90 & 22.43 $\pm$  3.50 & 0.669 $\pm$ 0.040 & 0.647 $\pm$ 0.061 & 0 & 0 &  OK  \\ 
OGLE-LMC-T2CEP-156 & WVir  &  15.39 & 33.84 $\pm$  3.31 & 0.631 $\pm$ 0.018 & 0.558 $\pm$ 0.021 & 1 & 0 &  OK  \\ 
OGLE-LMC-T2CEP-157 & WVir  &  14.33 & 27.71 $\pm$  1.38 & 0.439 $\pm$ 0.010 & 0.358 $\pm$ 0.010 & 0 & 0 &      \\ 
OGLE-LMC-T2CEP-158 & WVir  &   7.14 & 18.13 $\pm$  1.22 & 0.412 $\pm$ 0.010 & 0.337 $\pm$ 0.011 & 1 & 0 &      \\ 
OGLE-LMC-T2CEP-159 & WVir  &   6.63 & 18.90 $\pm$  0.47 & 0.478 $\pm$ 0.010 & 0.428 $\pm$ 0.010 & 0 & 0 &  ok  \\ 
OGLE-LMC-T2CEP-160 & BLHer &   1.76 &  9.15 $\pm$  1.08 & 0.500 $\pm$ 0.016 & 0.482 $\pm$ 0.023 & 0 & 0 &  OK  \\ 
OGLE-LMC-T2CEP-161 & WVir  &   8.53 & 29.73 $\pm$  3.39 & 1.004 $\pm$ 0.050 & 1.009 $\pm$ 0.085 & 0 & 0 &  OK  \\ 
OGLE-LMC-T2CEP-162 & RVTau &  30.39 & 44.46 $\pm$  3.27 & 0.516 $\pm$ 0.012 & 0.397 $\pm$ 0.012 & 1 & 0 &      \\ 
OGLE-LMC-T2CEP-163 & BLHer &   1.69 & 10.11 $\pm$  2.40 & 0.684 $\pm$ 0.085 & 0.676 $\pm$ 0.128 & 0 & 0 &  OK  \\ 
OGLE-LMC-T2CEP-164 & pWVir &   8.50 & 25.88 $\pm$  2.23 & 0.759 $\pm$ 0.019 & 0.688 $\pm$ 0.026 & 1 & 0 &  OK  \\ 
OGLE-LMC-T2CEP-165 & BLHer &   1.24 &  8.30 $\pm$  0.43 & 0.554 $\pm$ 0.011 & 0.647 $\pm$ 0.013 & 0 & 0 &  ok  \\ 
OGLE-LMC-T2CEP-166 & BLHer &   2.11 & 15.33 $\pm$  1.59 & 1.293 $\pm$ 0.070 & 1.518 $\pm$ 0.162 & 0 & 0 &  OK  \\ 
OGLE-LMC-T2CEP-167 & BLHer &   2.31 & 11.49 $\pm$  1.68 & 0.574 $\pm$ 0.027 & 0.586 $\pm$ 0.045 & 0 & 0 &  OK  \\ 
OGLE-LMC-T2CEP-168 & WVir  &  15.70 & 28.51 $\pm$  1.36 & 0.430 $\pm$ 0.010 & 0.334 $\pm$ 0.010 & 0 & 0 &      \\ 
OGLE-LMC-T2CEP-169 & RVTau &  30.96 & 37.17 $\pm$  7.37 & 0.377 $\pm$ 0.021 & 0.232 $\pm$ 0.015 & 1 & 0 &      \\ 
OGLE-LMC-T2CEP-170 & WVir  &   7.68 & 19.64 $\pm$  0.49 & 0.435 $\pm$ 0.010 & 0.375 $\pm$ 0.010 & 0 & 0 &      \\ 
OGLE-LMC-T2CEP-171 & BLHer &   1.55 &  8.56 $\pm$  1.35 & 0.523 $\pm$ 0.026 & 0.488 $\pm$ 0.036 & 0 & 0 &  OK  \\ 
OGLE-LMC-T2CEP-172 & WVir  &  11.22 & 25.68 $\pm$  2.57 & 0.492 $\pm$ 0.014 & 0.430 $\pm$ 0.016 & 0 & 0 &  OK  \\ 
OGLE-LMC-T2CEP-173 & WVir  &   4.15 & 16.30 $\pm$  0.99 & 0.523 $\pm$ 0.011 & 0.608 $\pm$ 0.014 & 0 & 0 &  ok  \\ 
OGLE-LMC-T2CEP-174 & RVTau &  46.82 & 75.02 $\pm$ 10.04 & 1.129 $\pm$ 0.084 & 0.848 $\pm$ 0.079 & 1 & 0 &  OK  \\ 
OGLE-LMC-T2CEP-175 & WVir  &   9.33 & 22.04 $\pm$  1.03 & 0.453 $\pm$ 0.010 & 0.378 $\pm$ 0.010 & 0 & 0 &      \\ 
OGLE-LMC-T2CEP-176 & WVir  &   7.99 & 20.49 $\pm$  0.93 & 0.470 $\pm$ 0.010 & 0.396 $\pm$ 0.010 & 0 & 0 &      \\ 
OGLE-LMC-T2CEP-177 & WVir  &  15.04 & 28.00 $\pm$  2.04 & 0.424 $\pm$ 0.011 & 0.341 $\pm$ 0.011 & 0 & 0 &      \\ 
OGLE-LMC-T2CEP-178 & WVir  &  12.21 & 23.84 $\pm$  1.74 & 0.378 $\pm$ 0.010 & 0.304 $\pm$ 0.011 & 0 & 0 &      \\ 
OGLE-LMC-T2CEP-179 & WVir  &   8.05 & 20.25 $\pm$  0.53 & 0.434 $\pm$ 0.010 & 0.379 $\pm$ 0.010 & 0 & 0 &  ok  \\ 
OGLE-LMC-T2CEP-180 & RVTau &  31.00 & 61.81 $\pm$ 10.03 & 1.147 $\pm$ 0.123 & 0.964 $\pm$ 0.142 & 1 & 0 &  OK  \\ 
OGLE-LMC-T2CEP-181 & pWVir &   7.21 & 23.87 $\pm$  2.63 & 0.762 $\pm$ 0.029 & 0.716 $\pm$ 0.041 & 0 & 0 &  OK  \\ 
OGLE-LMC-T2CEP-182 & WVir  &   8.23 & 23.35 $\pm$  1.11 & 0.606 $\pm$ 0.011 & 0.544 $\pm$ 0.011 & 0 & 0 &  ok  \\ 
OGLE-LMC-T2CEP-183 & WVir  &   6.51 & 18.78 $\pm$  1.11 & 0.455 $\pm$ 0.010 & 0.434 $\pm$ 0.011 & 0 & 0 &  OK  \\ 
OGLE-LMC-T2CEP-184 & WVir  &  14.84 & 27.43 $\pm$  2.21 & 0.393 $\pm$ 0.011 & 0.329 $\pm$ 0.011 & 0 & 0 &      \\ 
OGLE-LMC-T2CEP-185 & WVir  &  12.69 & 66.95 $\pm$  1.87 & 3.832 $\pm$ 0.057 & 5.143 $\pm$ 0.160 & 0 & 0 &      \\ 
OGLE-LMC-T2CEP-186 & WVir  &  16.36 & 31.99 $\pm$  1.66 & 0.513 $\pm$ 0.010 & 0.431 $\pm$ 0.011 & 0 & 0 &      \\ 
OGLE-LMC-T2CEP-187 & BLHer &   2.40 & 10.82 $\pm$  1.36 & 0.482 $\pm$ 0.017 & 0.465 $\pm$ 0.023 & 0 & 0 &  OK  \\ 
OGLE-LMC-T2CEP-188 & BLHer &   1.05 &  8.29 $\pm$  0.95 & 0.775 $\pm$ 0.031 & 0.843 $\pm$ 0.060 & 0 & 0 &  OK  \\ 
OGLE-LMC-T2CEP-189 & BLHer &   1.31 &  7.58 $\pm$  0.30 & 0.483 $\pm$ 0.010 & 0.459 $\pm$ 0.010 & 0 & 0 &  OK  \\ 
OGLE-LMC-T2CEP-190 & RVTau &  38.36 & 52.21 $\pm$  3.37 & 0.600 $\pm$ 0.012 & 0.425 $\pm$ 0.011 & 1 & 0 &      \\ 
OGLE-LMC-T2CEP-191 & RVTau &  34.34 & 63.59 $\pm$  5.29 & 1.105 $\pm$ 0.035 & 0.883 $\pm$ 0.038 & 1 & 0 &      \\ 
OGLE-LMC-T2CEP-192 & RVTau &  26.19 & 34.96 $\pm$  3.10 & 0.370 $\pm$ 0.011 & 0.257 $\pm$ 0.011 & 0 & 0 &      \\ 
OGLE-LMC-T2CEP-193 & WVir  &   7.00 & 20.00 $\pm$  0.49 & 0.516 $\pm$ 0.010 & 0.458 $\pm$ 0.010 & 0 & 0 &  ok  \\ 
OGLE-LMC-T2CEP-194 & BLHer &   1.31 &  8.25 $\pm$  0.64 & 0.592 $\pm$ 0.013 & 0.578 $\pm$ 0.017 & 0 & 0 &  OK  \\ 
OGLE-LMC-T2CEP-195 & BLHer &   2.75 & 11.49 $\pm$  1.15 & 0.485 $\pm$ 0.013 & 0.441 $\pm$ 0.016 & 0 & 0 &  OK  \\ 
OGLE-LMC-T2CEP-196 & WVir  &  14.96 & 32.84 $\pm$  2.33 & 0.627 $\pm$ 0.013 & 0.536 $\pm$ 0.014 & 0 & 0 &      \\ 
OGLE-LMC-T2CEP-197 & BLHer &   1.22 &  8.07 $\pm$  1.04 & 0.613 $\pm$ 0.025 & 0.608 $\pm$ 0.039 & 0 & 0 &  OK  \\ 
OGLE-LMC-T2CEP-198 & RVTau &  38.27 & 49.65 $\pm$  2.79 & 0.472 $\pm$ 0.010 & 0.373 $\pm$ 0.010 & 0 & 0 &      \\ 
OGLE-LMC-T2CEP-199 & RVTau &  37.20 & 38.00 $\pm$  2.67 & 0.377 $\pm$ 0.010 & 0.182 $\pm$ 0.010 & 1 & 0 &      \\ 
OGLE-LMC-T2CEP-200 & RVTau &  34.92 & 57.81 $\pm$  8.19 & 0.790 $\pm$ 0.047 & 0.661 $\pm$ 0.053 & 1 & 0 &  OK  \\ 
OGLE-LMC-T2CEP-201 & pWVir &  11.01 & 36.81 $\pm$  2.77 & 1.372 $\pm$ 0.044 & 1.207 $\pm$ 0.057 & 1 & 0 &  OK  \\ 
OGLE-LMC-T2CEP-202 & RVTau &  38.14 & 50.01 $\pm$  2.66 & 0.490 $\pm$ 0.010 & 0.382 $\pm$ 0.010 & 0 & 0 &      \\ 
OGLE-LMC-T2CEP-203 & RVTau &  37.13 & 46.46 $\pm$  3.67 & 0.427 $\pm$ 0.011 & 0.325 $\pm$ 0.011 & 0 & 0 &      \\ 
\hline
\end{tabular}
\end{table*}

\setcounter{table}{0}
\begin{table*}
\caption{Continued}

\begin{tabular}{lrrrrrrrrlrrrrr}
\hline
Name               & Type  & Period & Radius   & Mass$_{\rm Cep}$  & Mass$_{\rm RRL}$    & Dusty? & Binary? & Agree? \\ 
                   &       &  (d)   &  (\rsol) &     (\msol)     &     (\msol)       &        &         &            \\
\hline

OGLE-SMC-ACEP-01 & 1O    &   0.83 &  6.88 $\pm$  0.36 & 0.738 $\pm$ 0.012 & 0.728 $\pm$ 0.014 & 0 & 0 &  OK  \\ 
OGLE-SMC-ACEP-02 & F     &   0.83 &  7.87 $\pm$  0.17 & 0.923 $\pm$ 0.010 & 1.070 $\pm$ 0.011 & 0 & 0 &  ok  \\ 
OGLE-SMC-ACEP-03 & 1O    &   0.76 &  6.67 $\pm$  0.46 & 0.755 $\pm$ 0.015 & 0.765 $\pm$ 0.022 & 0 & 0 &  OK  \\ 
OGLE-SMC-ACEP-04 & F     &   0.83 &  8.23 $\pm$  0.32 & 1.046 $\pm$ 0.012 & 1.210 $\pm$ 0.019 & 0 & 0 &  ok  \\ 
OGLE-SMC-ACEP-05 & 1O    &   0.70 &  6.78 $\pm$  0.25 & 0.870 $\pm$ 0.011 & 0.928 $\pm$ 0.013 & 0 & 0 &  ok  \\ 
OGLE-SMC-ACEP-06 & F     &   1.26 &  9.12 $\pm$  0.68 & 0.824 $\pm$ 0.018 & 0.820 $\pm$ 0.027 & 0 & 0 &  OK  \\ 
OGLE-SMC-T2CEP-001 & pWVir &  11.87 & 43.32 $\pm$  1.89 & 1.785 $\pm$ 0.029 & 1.685 $\pm$ 0.040 & 0 & 0 &  OK  \\ 
OGLE-SMC-T2CEP-002 & BLHer &   1.37 &  7.64 $\pm$  0.67 & 0.484 $\pm$ 0.012 & 0.434 $\pm$ 0.014 & 0 & 0 &  OK  \\ 
OGLE-SMC-T2CEP-003 & WVir  &   4.36 & 12.73 $\pm$  1.28 & 0.347 $\pm$ 0.011 & 0.278 $\pm$ 0.011 & 0 & 0 &      \\ 
OGLE-SMC-T2CEP-004 & WVir  &   6.53 & 19.98 $\pm$  2.56 & 0.569 $\pm$ 0.022 & 0.511 $\pm$ 0.028 & 0 & 0 &  OK  \\ 
OGLE-SMC-T2CEP-005 & WVir  &   8.21 & 19.41 $\pm$  0.45 & 0.400 $\pm$ 0.010 & 0.326 $\pm$ 0.010 & 0 & 0 &      \\ 
OGLE-SMC-T2CEP-006 & BLHer &   1.24 &  7.33 $\pm$  0.15 & 0.485 $\pm$ 0.010 & 0.459 $\pm$ 0.010 & 0 & 0 &  OK  \\ 
OGLE-SMC-T2CEP-007 & RVTau &  30.96 & 77.35 $\pm$ 18.57 & 2.058 $\pm$ 0.750 & 1.804 $\pm$ 0.865 & 0 & 1 &  OK  \\ 
OGLE-SMC-T2CEP-008 & BLHer &   1.49 & 10.45 $\pm$  0.44 & 0.838 $\pm$ 0.011 & 0.913 $\pm$ 0.015 & 0 & 0 &  ok  \\ 
OGLE-SMC-T2CEP-009 & BLHer &   2.97 & 12.97 $\pm$  0.58 & 0.572 $\pm$ 0.010 & 0.547 $\pm$ 0.011 & 0 & 0 &  OK  \\ 
OGLE-SMC-T2CEP-010 & pWVir &  17.48 & 60.53 $\pm$  6.25 & 2.347 $\pm$ 0.226 & 2.296 $\pm$ 0.360 & 0 & 1 &  OK  \\ 
OGLE-SMC-T2CEP-011 & pWVir &   9.93 & 38.92 $\pm$  2.65 & 1.891 $\pm$ 0.068 & 1.665 $\pm$ 0.089 & 1 & 0 &  OK  \\ 
OGLE-SMC-T2CEP-012 & RVTau &  29.22 & 41.67 $\pm$  1.15 & 0.486 $\pm$ 0.010 & 0.352 $\pm$ 0.010 & 0 & 0 &      \\ 
OGLE-SMC-T2CEP-013 & WVir  &  13.81 & 25.27 $\pm$  0.67 & 0.386 $\pm$ 0.010 & 0.293 $\pm$ 0.010 & 0 & 0 &      \\ 
OGLE-SMC-T2CEP-014 & WVir  &  13.88 & 23.85 $\pm$  0.59 & 0.335 $\pm$ 0.010 & 0.247 $\pm$ 0.010 & 0 & 0 &      \\ 
OGLE-SMC-T2CEP-015 & BLHer &   2.57 & 12.70 $\pm$  1.04 & 0.767 $\pm$ 0.018 & 0.649 $\pm$ 0.021 & 0 & 0 &      \\ 
OGLE-SMC-T2CEP-016 & BLHer &   2.11 & 10.22 $\pm$  0.63 & 0.520 $\pm$ 0.011 & 0.487 $\pm$ 0.012 & 0 & 0 &  OK  \\ 
OGLE-SMC-T2CEP-017 & BLHer &   1.30 &  8.79 $\pm$  0.97 & 0.697 $\pm$ 0.024 & 0.703 $\pm$ 0.040 & 0 & 0 &  OK  \\ 
OGLE-SMC-T2CEP-018 & RVTau &  39.52 & 57.52 $\pm$  6.83 & 0.742 $\pm$ 0.031 & 0.531 $\pm$ 0.027 & 1 & 0 &      \\ 
OGLE-SMC-T2CEP-019 & RVTau &  40.91 & 50.40 $\pm$  2.09 & 0.541 $\pm$ 0.010 & 0.347 $\pm$ 0.010 & 1 & 0 &      \\ 
OGLE-SMC-T2CEP-020 & RVTau &  50.62 & 50.15 $\pm$  3.50 & 0.376 $\pm$ 0.010 & 0.243 $\pm$ 0.010 & 0 & 0 &      \\ 
OGLE-SMC-T2CEP-021 & BLHer &   2.31 &  8.99 $\pm$  0.37 & 0.344 $\pm$ 0.010 & 0.294 $\pm$ 0.010 & 0 & 0 &      \\ 
OGLE-SMC-T2CEP-022 & BLHer &   1.47 &  7.22 $\pm$  0.57 & 0.364 $\pm$ 0.011 & 0.331 $\pm$ 0.011 & 0 & 0 &  OK  \\ 
OGLE-SMC-T2CEP-023 & pWVir &  17.68 & 39.21 $\pm$  1.69 & 0.820 $\pm$ 0.011 & 0.670 $\pm$ 0.012 & 0 & 1 &      \\ 
OGLE-SMC-T2CEP-024 & RVTau &  43.96 & 51.41 $\pm$  3.24 & 0.506 $\pm$ 0.011 & 0.326 $\pm$ 0.010 & 1 & 0 &      \\ 
OGLE-SMC-T2CEP-025 & pWVir &  14.17 & 30.97 $\pm$  1.42 & 0.624 $\pm$ 0.011 & 0.496 $\pm$ 0.011 & 0 & 1 &      \\ 
OGLE-SMC-T2CEP-026 & BLHer &   1.70 & 10.17 $\pm$  0.39 & 0.696 $\pm$ 0.011 & 0.680 $\pm$ 0.011 & 0 & 0 &  OK  \\ 
OGLE-SMC-T2CEP-027 & BLHer &   1.54 &  7.70 $\pm$  0.31 & 0.408 $\pm$ 0.010 & 0.368 $\pm$ 0.010 & 0 & 0 &  OK  \\ 
OGLE-SMC-T2CEP-028 & pWVir &  15.26 & 49.73 $\pm$  1.33 & 1.653 $\pm$ 0.014 & 1.655 $\pm$ 0.018 & 0 & 1 &  OK  \\ 
OGLE-SMC-T2CEP-029 & RVTau &  33.68 & 91.49 $\pm$  2.38 & 2.547 $\pm$ 0.025 & 2.524 $\pm$ 0.035 & 0 & 1 &  OK  \\ 
OGLE-SMC-T2CEP-030 & BLHer &   3.39 & 14.97 $\pm$  0.81 & 0.750 $\pm$ 0.012 & 0.658 $\pm$ 0.014 & 0 & 0 &  ok  \\ 
OGLE-SMC-T2CEP-031 & WVir  &   7.90 & 20.76 $\pm$  2.32 & 0.491 $\pm$ 0.015 & 0.419 $\pm$ 0.017 & 0 & 0 &      \\ 
OGLE-SMC-T2CEP-032 & WVir  &  14.25 & 26.57 $\pm$  1.75 & 0.434 $\pm$ 0.011 & 0.320 $\pm$ 0.010 & 1 & 0 &      \\ 
OGLE-SMC-T2CEP-033 & BLHer &   1.88 & 11.56 $\pm$  0.67 & 0.831 $\pm$ 0.014 & 0.831 $\pm$ 0.019 & 0 & 0 &  OK  \\ 
OGLE-SMC-T2CEP-034 & WVir  &  20.12 & 36.42 $\pm$  1.09 & 0.566 $\pm$ 0.010 & 0.442 $\pm$ 0.010 & 0 & 0 &      \\ 
OGLE-SMC-T2CEP-035 & WVir  &  17.18 & 29.80 $\pm$  3.18 & 0.437 $\pm$ 0.013 & 0.326 $\pm$ 0.013 & 1 & 0 &      \\ 
OGLE-SMC-T2CEP-036 & BLHer &   1.09 &  9.38 $\pm$  0.36 & 1.018 $\pm$ 0.012 & 1.117 $\pm$ 0.017 & 0 & 0 &  ok  \\ 
OGLE-SMC-T2CEP-037 & BLHer &   1.56 & 10.03 $\pm$  0.41 & 0.737 $\pm$ 0.011 & 0.757 $\pm$ 0.012 & 0 & 0 &  OK  \\ 
OGLE-SMC-T2CEP-038 & pWVir &   4.44 & 21.40 $\pm$  1.62 & 1.203 $\pm$ 0.035 & 1.152 $\pm$ 0.053 & 0 & 0 &  OK  \\ 
OGLE-SMC-T2CEP-039 & BLHer &   1.89 & 11.52 $\pm$  0.25 & 0.783 $\pm$ 0.010 & 0.818 $\pm$ 0.010 & 0 & 0 &  OK  \\ 
OGLE-SMC-T2CEP-040 & WVir  &  16.11 & 29.73 $\pm$  3.47 & 0.465 $\pm$ 0.015 & 0.360 $\pm$ 0.015 & 0 & 0 &      \\ 
OGLE-SMC-T2CEP-041 & RVTau &  29.12 & 37.67 $\pm$  1.72 & 0.400 $\pm$ 0.010 & 0.267 $\pm$ 0.010 & 0 & 0 &      \\ 
OGLE-SMC-T2CEP-042 & BLHer &   1.49 &  8.54 $\pm$  0.18 & 0.544 $\pm$ 0.010 & 0.521 $\pm$ 0.010 & 0 & 0 &  OK  \\ 
OGLE-SMC-T2CEP-043 & RVTau &  23.74 & 41.40 $\pm$  2.07 & 0.620 $\pm$ 0.011 & 0.484 $\pm$ 0.011 & 0 & 0 &      \\ 
\hline
\end{tabular}
\end{table*}

\end{appendix}

\end{document}